\documentclass[11pt,twoside]{article}
\usepackage{amsmath,amssymb,latexsym,graphicx,hyperref,epstopdf,subcaption,color}
\usepackage{algpseudocode,algorithm,epstopdf}
\usepackage[capitalise]{cleveref}
\normalsize
\newtheorem{theorem}{Theorem}
\newtheorem{lemma}{Lemma}

\newtheorem{fact}{Fact}

\newenvironment{proof}{{\sc Proof. }}{\hfill$\Box$\vspace{0.2in}}

\def\mcC{\mathcal{C}}
\def\mcI{\mathcal{I}}
\def\mcP{\mathcal{P}}
\def\mcQ{\mathcal{Q}}

\title{Approximation algorithms for the directed path partition problems}
\author{
	Yong~Chen\thanks{\texttt{chenyong,anzhang@hdu.edu.cn}.
	Department of Mathematics, Hangzhou Dianzi University.  Hangzhou, China.}
\and
	Zhi-Zhong~Chen\thanks{\texttt{zzchen@mail.dendai.ac.jp}.
	Division of Information System Design, Tokyo Denki University.  Saitama, Japan.}
\and
	Curtis~Kennedy\thanks{\texttt{ckennedy,guohui@ualberta.ca}.
	Department of Computing Science, University of Alberta.  Edmonton, Alberta T6G 2E8, Canada.}
\and
	Guohui~Lin$^\ddagger$\thanks{Correspondence author.}
\and
	Yao~Xu\thanks{\texttt{yxu@georgiasouthern.edu}.
	Department of Computer Science, Georgia Southern University. Statesboro, USA.}
\and
	An Zhang$^*$}%
\date{\today}
\begin{document}
\maketitle

\begin{abstract}
Given a directed graph $G = (V, E)$, the $k$-path partition problem is to find a minimum collection of vertex-disjoint directed paths
each of order at most $k$ to cover all the vertices of $V$.
The problem has various applications in facility location, network monitoring, transportation and others.
Its special case on undirected graphs has received much attention recently, but the general directed version is seemingly untouched in the literature.
We present the first $k/2$-approximation algorithm, for any $k \ge 3$,
based on a novel concept of augmenting path to minimize the number of singletons in the partition.
When $k \ge 7$, we present an improved $(k+2)/3$-approximation algorithm based on the maximum path-cycle cover followed by
a careful $2$-cycle elimination process.
When $k = 3$, we define the second novel kind of augmenting paths and propose an improved $13/9$-approximation algorithm.

\end{abstract}

\subsubsection*{Acknowledgements.}
This research is supported by the NSFC Grants 11771114 and 11971139 (YC and AZ),
the Zhejiang Provincial NSFC Grant LY21A010014 (YC and AZ),
the CSC Grants 201508330054 (YC) and 201908330090 (AZ),
the Grant-in-Aid for Scientific Research of the Ministry of Education, Science, Sports and Culture of Japan Grant No. 18K11183 (ZZC),
and the NSERC Canada (GL).

\newpage
\section{Introduction}
Given a directed graph $G$, we consider the simple directed paths in the graph.
We denote the vertex set and the edge set of $G$ by $V(G)$ and $E(G)$, respectively, and simplify them as $V$ and $E$ when $G$ is clear from the context,
that is, $G = (V, E)$.
We assume without loss of generality that there are no self-loops or multiple edges in the graph.
Let $n = |V|$ and $m = |E|$, which are referred to as the {\em order} and the {\em size} of the graph $G$, respectively.
For a vertex $v$ in $G$, the number of edges entering (leaving, respectively) $v$ is denoted by $d_G^-(v)$ ($d_G^+(v)$, respectively),
which is referred to as the {\em in-degree} ({\em out-degree}, respectively) of $v$. 
As well, $d_G^-(v)$ and $d_G^+(v)$ are simplified as $d^-(v)$ and $d^+(v)$, respectively, when the graph $G$ is clear from the context.
We note that an undirected edge is deemed bidirectional; this way, an undirected graph is a special directed graph.

A {\em simple directed path} in the graph is a sequence of distinct vertices so that there is an edge from every vertex to its succeeding one.
For convenience, these edges are said to be the edges of the path.
When there is an edge from the last vertex to the first vertex, then adding this edge to the path gives rise to a {\em simple directed cycle}.
In the sequel, we leave out both ``simple'' and ``directed'', and simply call them a path and a cycle, respectively.

The {\em order} ({\em length}, respectively) of a path is the number of vertices (edges, respectively) on the path and
an order-$k$ path is simply called a {\em $k$-path} (or sometimes, a length-$(k-1)$ path).
The {\em $k$-path partition} (abbreviated as $k$PP) problem is to find a minimum collection of vertex-disjoint paths each of order at most $k$
such that every vertex is on some path in the collection.

When $k$ is part of the input, the $k$PP problem includes the NP-complete {\sc Hamiltonian Path} problem~\cite{GJ79} as a special case,
and thus it is APX-hard and is not approximable within $2$ unless P = NP.
On the other hand, the $2$PP problem is equivalent to the {\sc Maximum Matching} problem on undirected graphs (by ignoring the edge directions, if any),
which is solvable in $O(m \sqrt{n} \log(n^2/m)/\log n)$-time~\cite{GK04}.
In the sequel, we assume $k \ge 3$ is a fixed constant.
The $k$PP problem on undirected graphs has received a number of studies~\cite{YCH97,MT07,CGL19a,CGL19b,CGS19}.
For example, the problem is solvable in polynomial time for trees~\cite{YCH97}, cographs~\cite{Ste00} and bipartite permutation graphs~\cite{Ste03}.
When $k = 3$, that is, for $3$PP, Monnot and Toulouse~\cite{MT07} presented a $3/2$-approximation algorithm;
the approximation ratio has been improved to $13/9$~\cite{CGL19a}, $4/3$~\cite{CGL19b} and the current best $21/16$~\cite{CGS19}.
For any fixed $k \ge 3$, $k$PP on undirected graphs admits a $k/2$-approximation algorithm~\cite{CGL19b}.
The intractability of $k$PP, when $k$ is part of the input or a fixed constant,
on some special undirected graph classes (such as chordal, bipartite, comparability, and cographs)
has been investigated and depicted~\cite{Ste00,Ste03,Kor18}.

It is noted that in various applications such as facility location, network monitoring, and transportation,
the background network is modeled as a directed graph.
However, the general $k$PP problem on directed graphs is seemingly untouched in the literature.
One sees that the $k$PP problem can be regarded as a special case of the minimum {\sc Exact $k$-Set Cover},
by creating a subset of $\ell$ vertices, for all $\ell \le k$, if and only if they are traceable in the input graph.
The {\sc Exact $k$-Set Cover} problem is one of Karp's $21$ NP-complete problems~\cite{Kar72},
and its minimization variant does not admit any non-trivial approximation algorithms.

In this paper, we investigate the $k$PP problem on directed graphs from the approximation algorithm perspective.
We present the first $k/2$-approximation algorithm, for any $k \ge 3$,
based on a novel concept of augmenting path to minimize the number of singletons in the partition.
When $k \ge 7$, we present an improved $(k+2)/3$-approximation algorithm based on the maximum path-cycle cover followed by
a careful $2$-cycle elimination process.
Lastly, for $3$PP, we define the second novel kind of augmenting paths and propose an improved $13/9$-approximation algorithm.
The state-of-the-art approximation results for the $k$PP problems are summarized in Table~\ref{tab01}.

\begin{table}[htb]
\caption{The best known approximation ratios for the $k$PP problems;
	those labeled with $*$ are achieved in this paper.\label{tab01}}
\begin{center}
\begin{tabular}{r|l||l}
				&$k \ge 3$ fixed							&$k/2$-approx$^*$\\
				\cline{2-3}
directed	&$k \ge 7$ fixed							&$(k+2)/3$-approx$^*$\\
				\cline{2-3}
				&$k = 3$										&$13/9$-approx$^*$\\
\hline
\hline
				&$k \ge 3$ fixed							&$k/2$-approx~\cite{CGL19b} ($k = 3$~\cite{MT07})\\
				\cline{2-3}
undirected	&$k \ge 7$ fixed							&$(k+2)/3$-approx$^*$\\
				\cline{2-3}
				&$k = 3$										&$21/16$-approx~\cite{CGS19}\\
\end{tabular}
\end{center}
\end{table}

The rest of the paper is organized as follows:
In Section~\ref{sec2} we present our three approximation algorithms, each in a separate subsection.
For the last $13/9$-approximation algorithm for $3$PP, we also provide a series of instances to show the tightness of the approximation ratio.
We conclude the paper in Section~\ref{sec3}, with several future work.

\section{Approximation algorithms}
\label{sec2}
Given a directed graph $G = (V, E)$ and a positive integer $b$,
a {\em $b$-matching} $M$ is a subset of edges so that there are at most $b$ of them entering every vertex of $V$ and
at most $b$ of them leaving every vertex of $V$.
One sees that a $1$-matching in the directed graph $G = (V, E)$ consists of vertex-disjoint paths and cycles,
and thus it is also called a {\em path-cycle cover}.
A maximum path-cycle cover of the graph $G$ can be computed in $O(m n \log n)$ time~\cite{Gab83}.

We want to remind the readers that a $b$-matching can be defined in the same way for an undirected graph,
where the edges are deemed bidirectional.
Therefore, a $2$-matching in the undirected graph consists of vertex-disjoint paths and cycles, which is also called a path-cycle cover.
A $1$-matching in the undirected graph is simply called a matching.
With respect to a matching $M$, an edge of $M$ is called a {\em matched} edge, or otherwise a {\em free} edge;
a vertex incident with a matched edge is said matched, or otherwise free.
Furthermore, an {\em alternating} path is one with alternating free and matched edges.
An {\em augmenting} path is an alternating path that begins and ends with free vertices.
An augmenting path is used to increase the size of the matching $M$, through replacing the matched edges on the path by the free edges on the path.

With respect to a $k$-path partition in a directed graph,
below we will define what the matched edges and the free edges are, and two novel kinds of alternating and augmenting paths.
The augmenting paths are used to improve the $k$-path partition (that is, to increase the number of edges in the $k$-path partition).

In Section~\ref{sec21}, we present a first $k/2$-approximation for $k$PP for any $k \ge 3$,
in which a key ingredient is the first novel kind of alternating and augmenting paths to minimize the number of {\em singletons} in a $k$-path partition.
We realize that most argument in this subsection is a nontrivial generalization of the undirected counterpart.
We therefore present only the design of the algorithm, while leaving the detailed analysis to Appendix~\ref{secA}.
Section~\ref{sec22} deals with the case where $k \ge 7$;
we start with a maximum path-cycle cover, to design a $(k+2)/3$-approximation algorithm by carefully dealing with $2$-cycles in the cover.
Lastly in Section~\ref{sec23}, we design a $13/9$-approximation algorithm for $3$PP,
in which a key ingredient is the second novel kind of alternating and augmenting paths to reduce the number of $2$-paths in a $k$-path partition.
We also show that the approximation ratio $13/9$ is tight for the algorithm.
We point out that the structural properties of a $3$-path partition differ much for directed graphs and for undirected graphs,
for example, the maximum matching based $3/2$-approximation algorithm by Monnot and Toulouse~\cite{MT07} and
the local search based approximation algorithms~\cite{CGL19a,CGL19b,CGS19} for $3$PP on undirected graphs do not extend to the directed graphs.

\subsection{A first $k/2$-approximation for $k$PP}\label{sec21}
Suppose we are given a directed graph $G = (V, E)$ and a $k$-path partition $\mcQ$ of $G$.

For ease of presentation, the edges on the paths of $\mcQ$ are simply called the edges of $\mcQ$,
and the $1$-paths of $\mcQ$ are called {\em singletons} of $\mcQ$.
For an $\ell$-path $v_1$-$v_2$-$\cdots$-$v_\ell$ in $\mcQ$ (where $\ell \le k$), $v_j$ is called the $j$-th vertex on the path,
and in particular $v_1$ is the head vertex and $v_\ell$ is the tail vertex.
The intention of a to-be-defined augmenting path is to reduce the number of singletons, through adding an edge, so as to improve $\mcQ$.

Let us first define the two types of edges that can be on alternating paths.
Each alternating path starts with a singleton.
If there is no singleton in the $k$-path partition $\mcQ$,
then we do not bother to define the edge types and $\mcQ$ is our {\em desired solution} achieving the minimum number of singletons.
In the other case, for an $\ell$-path $v_1$-$v_2$-$\cdots$-$v_\ell$ in $\mcQ$,
the edges $(v_1, v_2)$ and $(v_{\ell-1}, v_\ell)$ are {\em matched} edges;
all the edges incident at $v_1$ ($v_\ell$, respectively) in the graph $G = (V, E)$,
both entering and leaving $v_1$ ($v_\ell$, respectively), except $(v_1, v_2)$ ($(v_{\ell-1}, v_\ell)$, respectively)
are {\em free} edges with respect to $\mcQ$.
We note that there are edges which are neither matched nor free, such as the edge $(v_2, v_3)$ when $\ell \ge 4$;
we may call them {\em irrelevant} edges.

\begin{figure}[ht]
\centering
\begin{subfigure}{0.45\textwidth}
  \setlength{\unitlength}{0.7bp}%
  \begin{picture}(206.01, 90.15)(0,0)
  \put(0,0){\includegraphics[scale=0.7]{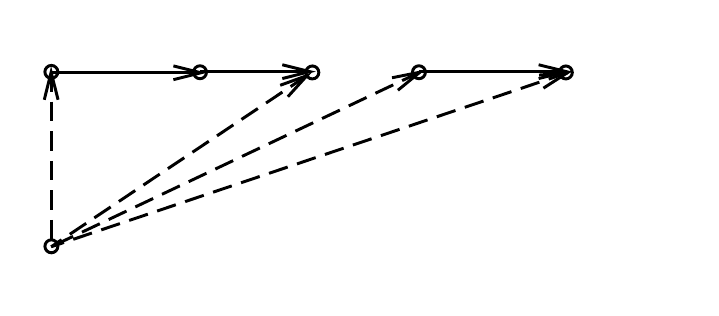}}
  \put(9.46,75.48){\fontsize{11.38}{13.66}\selectfont $v_1$}
  \put(5.67,8.12){\fontsize{11.38}{13.66}\selectfont $u$}
  \put(54.43,75.59){\fontsize{11.38}{13.66}\selectfont $v_2$}
  \put(86.84,75.42){\fontsize{11.38}{13.66}\selectfont $v_3$}
  \put(163.17,75.42){\fontsize{11.38}{13.66}\selectfont $v_\ell$}
  \put(97.37,68.68){\fontsize{11.38}{13.66}\selectfont $\ldots$}
  \end{picture}%
\caption{\label{fig01a}}
\end{subfigure}
\begin{subfigure}{0.45\textwidth}
  \setlength{\unitlength}{0.7bp}%
  \begin{picture}(206.00, 90.15)(0,0)
  \put(0,0){\includegraphics[scale=0.7]{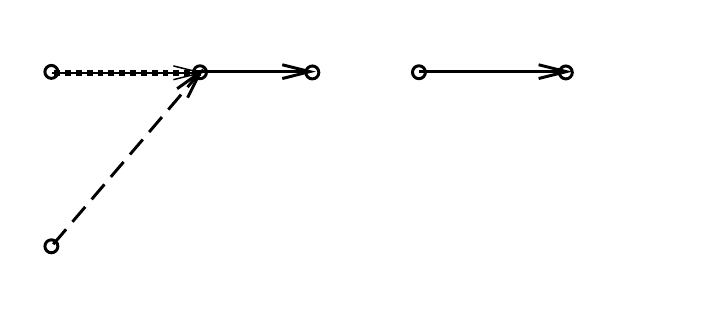}}
  \put(9.46,75.48){\fontsize{11.38}{13.66}\selectfont $v_1$}
  \put(5.67,8.12){\fontsize{11.38}{13.66}\selectfont $u$}
  \put(54.43,75.59){\fontsize{11.38}{13.66}\selectfont $v_2$}
  \put(86.84,75.42){\fontsize{11.38}{13.66}\selectfont $v_3$}
  \put(163.17,75.42){\fontsize{11.38}{13.66}\selectfont $v_\ell$}
  \put(97.37,68.68){\fontsize{11.38}{13.66}\selectfont $\ldots$}
  \end{picture}%
\caption{\label{fig01b}}
\end{subfigure}
\begin{subfigure}{0.45\textwidth}
  \setlength{\unitlength}{0.7bp}%
  \begin{picture}(197.18, 90.15)(0,0)
  \put(0,0){\includegraphics[scale=0.7]{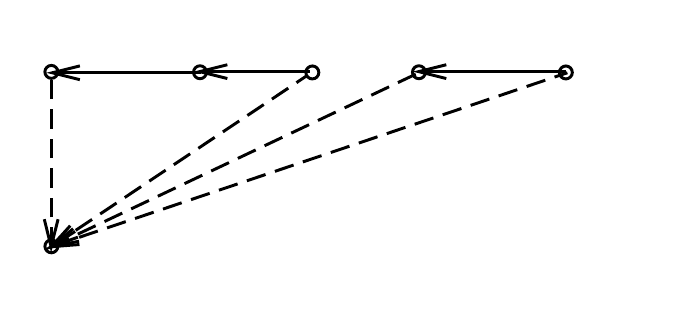}}
  \put(9.46,75.48){\fontsize{11.38}{13.66}\selectfont $v_\ell$}
  \put(5.67,8.12){\fontsize{11.38}{13.66}\selectfont $u$}
  \put(54.43,75.59){\fontsize{11.38}{13.66}\selectfont $v_{\ell-1}$}
  \put(86.84,75.42){\fontsize{11.38}{13.66}\selectfont $v_{\ell-2}$}
  \put(163.17,75.42){\fontsize{11.38}{13.66}\selectfont $v_1$}
  \put(90.16,65.47){\fontsize{11.38}{13.66}\selectfont $\ldots$}
  \end{picture}%
\caption{\label{fig01c}}
\end{subfigure}
\begin{subfigure}{0.45\textwidth}
  \setlength{\unitlength}{0.7bp}%
  \begin{picture}(197.18, 90.15)(0,0)
  \put(0,0){\includegraphics[scale=0.7]{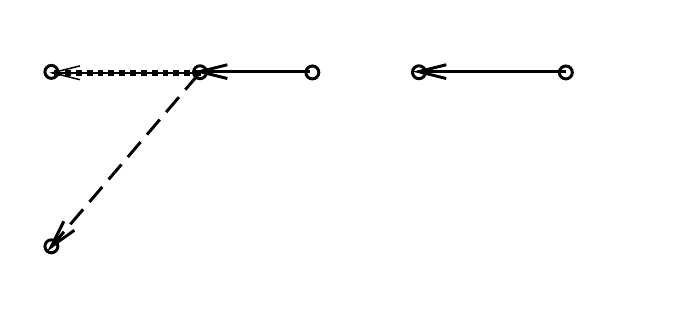}}
  \put(9.46,75.48){\fontsize{11.38}{13.66}\selectfont $v_\ell$}
  \put(5.67,8.12){\fontsize{11.38}{13.66}\selectfont $u$}
  \put(54.43,75.59){\fontsize{11.38}{13.66}\selectfont $v_{\ell-1}$}
  \put(86.84,75.42){\fontsize{11.38}{13.66}\selectfont $v_{\ell-2}$}
  \put(163.17,75.42){\fontsize{11.38}{13.66}\selectfont $v_1$}
  \put(93.37,68.68){\fontsize{11.38}{13.66}\selectfont $\ldots$}
  \end{picture}%
\caption{\label{fig01d}}
\end{subfigure}
\caption{(a) and (b), a free edge $(u, v)$ for all possible configurations of the vertex $v$ in the $k$-path partition $\mcQ$,
	where the dashed edges are free, the solid edges are in $\mcQ$, and the additionally dotted edge is matched.
	(c) and (d) show the symmetric case for a free edge $(v, u)$.
	An alternating path is one alternating free and matched edges, and it starts with a singleton.\label{fig01}}
\end{figure}
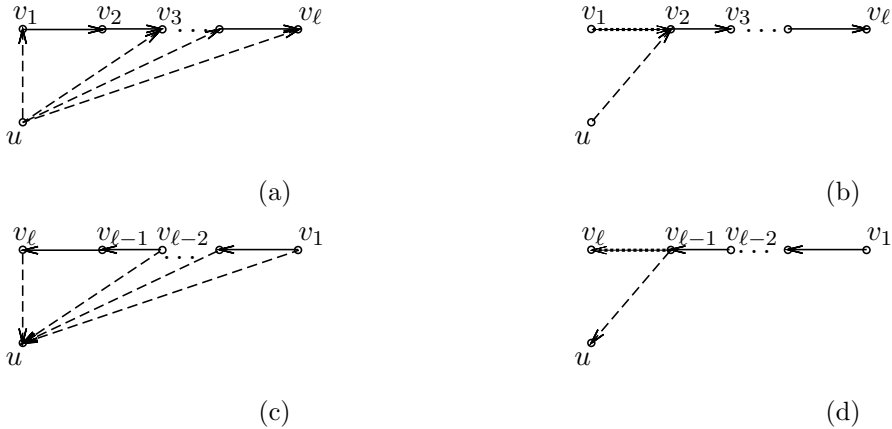

Consider a singleton $u$ and a free edge $(u, v)$.
(The following argument applies to the symmetric case $(v, u)$ by reversing the direction of the involved edges, if any.)
If there is no matched edge entering $v$, that is, $v$ is not the second vertex of any path of $\mcQ$ (see Figure~\ref{fig01a} for illustrations),
then the alternating path ends.
In Lemma~\ref{lemma09}, we will show that such an alternating path is an augmenting path,
and it can be used to transfer $\mcQ$ into another $k$-path partition
(by taking the {\em symmetric difference} of $\mcQ$ and the augmenting path) with at least one less singleton.

If there is a matched edge entering $v$, that is, $v$ is the second vertex $v_2$ of an $\ell$-path $v_1$-$v_2$-$\cdots$-$v_\ell$ of $\mcQ$,
but the matched edge $(v_1, v_2)$ has already been included in the alternating path,
then the alternating path ends too and it is not an augmenting path.

In the other case where $v$ is the second vertex $v_2$ of an $\ell$-path $v_1$-$v_2$-$\cdots$-$v_\ell$ of $\mcQ$
(see Figure~\ref{fig01b} for an illustration),
the matched edge $(v_1, v_2)$, which is shown as the dotted edge in Figure~\ref{fig01b}, extends the alternating to the vertex $v_1$.
Iteratively, when there is no free edge incident at $v_1$ and outside of the alternating path, the alternating path ends;
or otherwise one such free edge extends the alternating path and we may repeat the above process on this newly added free edge.
Therefore, either a free edge or a matched edge ends the alternating path,
and an augmenting path is achieved if and only if a free edge ends the alternating path so that there is no matched edge incident at the last vertex.
In the Appendix, we show in Lemma~\ref{lemma08} that the entire process can be done via a breadth-first-search (BFS) traversal and takes $O(m)$ time;
and show in Lemma~\ref{lemma09} that an augmenting path can be used to transfer $\mcQ$ into another $k$-path partition
with at least one less singleton in $O(n)$ time.

\begin{figure}[htb]
\begin{center}
\framebox{
\begin{minipage}{5.5in}
Algorithm {\sc Approx1}:\\
Input: a directed graph $G = (V, E)$;\\
Output: a $k$-path partition $\mcQ$.
\begin{itemize}
\parskip=0pt
\item[1.]
	$\mcQ$ is initialized to contain $n$ singletons.
\item[2.]
	For each singleton $u$ in $\mcQ$,
	\begin{itemize}
	\parskip=0pt
	\item[2.1]
		explore the alternating paths starting with $u$ via a BFS traversal;
	\item[2.2]
		if an augmenting path is found, update $\mcQ$ and break to restart Step 2.
	\end{itemize}
\item[3.]
	Return $\mcQ$.
\end{itemize}
\end{minipage}}
\end{center}
\caption{A high level description of the algorithm {\sc Approx1}.\label{fig02}}
\end{figure}

We now describe our algorithm {\sc Approx1} for computing a $k$-path partition $\mcQ$ in a directed graph $G = (V, E)$.
The algorithm is iterative,
and in each iteration it tries to find an augmenting path starting with a singleton in the current $k$-path partition,
and uses it to transfer into another $k$-path partition with at least one less singleton.
To this purpose,
the initial $k$-path partition is set to contain $n$ singletons;
in each iteration, the algorithm explores all the alternating paths each starting with a singleton.
The iteration terminates at any time when an augmenting path is found, followed by updating the $k$-path partition;
if no augmenting path is found for any singleton, then the algorithm terminates and returns the current $k$-path partition as the solution.
From Lemmas~\ref{lemma08} and \ref{lemma09}, the overall running time is $O(nm)$.
A high level description of the algorithm is depicted in Figure~\ref{fig02}.

\begin{theorem}
\label{thm01}
The algorithm {\sc Approx1} is an $O(nm)$-time $k/2$-approximation for the $k$PP problem.
\end{theorem}
\begin{proof}
See Appendix.
\end{proof}

\subsection{An improved $(k+2)/3$-approximation for $k$PP, when $k \ge 7$}\label{sec22}
We fix an integer $k \ge 7$. 
Given a directed graph $G = (V, E)$,
our $(k+2)/3$-approximation algorithm {\sc Approx2} for $k$PP starts by performing the first three steps as in Figure~\ref{fig03},
in which a maximum path-cycle cover $\mcC$ of $G$ is computed and a subgraph $G_1$ of $G$ is constructed. 

\begin{figure}[htb]
\begin{center}
\framebox{
\begin{minipage}{5.5in}
Algorithm {\sc Approx2}:\\
Input: a directed graph $G$;\\
Output: a $k$-path partition.
\begin{enumerate}
\parskip=0pt
\item\label{step:H} 
	Compute a maximum path-cycle cover $\mcC$ of $G$.
\item\label{step:modify} 
	While $\exists (u, v) \in E(G)-E(\mcC)$ such that $d_\mcC^+(u) = 0$ (respectively, $d_\mcC^-(v) = 0$) 
	and $v$ (respectively, $u$) is on some cycle $C$ of $\mcC$,
	\begin{itemize}
	\parskip=0pt
	\item[2.1]
		replace the edge entering $v$ (respectively, leaving $u$) in $\mcC$ by $(u,v)$. 
	\end{itemize}
\item\label{step:G1} 
	Construct a directed graph $G_1 = (V(G), E_1)$,
	where $E_1$ is the set of all edges $(u, v) \in E(G) - E(\mcC)$ such that 
	$u$ and $v$ appear in different connected components of $\mcC$, at least one of which is a $2$-cycle.
\end{enumerate} 
\end{minipage}}
\end{center}
\caption{The description of the algorithm {\sc Approx2}, the first three steps (to be continued).\label{fig03}}
\end{figure}

Hereafter, $\mcC$ always refers to the path-cycle cover obtained after the completion of Step~\ref{step:modify}.
We give several definitions related to the graphs $\mcC$ and $G_1$.
A {\em path} (respectively, {\em cycle}) {\em component} of $\mcC$ is a connected component of $\mcC$ that is a path (respectively, cycle).
For convenience, we say that two connected components $C_1$ and $C_2$ of $\mcC$ are {\em adjacent} in a subgraph $G'$ of $G$
if there is an edge $(u_1, u_2) \in E(G')$ such that $u_1 \in V(C_1)$ and $u_2 \in V(C_2)$. 
Let $S$ be a subgraph of $G_1$. 
$S$ {\em saturates} a $2$-cycle $C$ of $\mcC$ if at least one edge of $S$ is incident at a vertex of $C$. 
The {\em weight} of $S$ is defined as the number of $2$-cycles of $\mcC$ saturated by $S$.

\begin{lemma}\label{lem:2match}
A maximum-weighted path-cycle cover in $G_1$ can be computed in $O(nm\log n)$ time. 
\end{lemma}
\begin{proof} 
The proof is done by a reduction to the maximum-weight $[f,g]$-factor problem,
which is known to be solvable in $O(n' m' \log n')$ time~\cite{Gab83} for a given edge-weighted undirected graph ${\cal G}$ with $n'$ vertices and $m'$ edges.

Recall that for two functions $f$ and $g$ mapping each vertex $v$ of the graph ${\cal G}$ to an integer with $f(v) \le g(v)$,
an [$f,g$]{\em -factor} of ${\cal G}$ is a subgraph ${\cal H}$ of ${\cal G}$ such that $V({\cal H}) = V({\cal G})$ and 
$f(v) \le d_{\cal H}(v) \le g(v)$ for every $v \in V({\cal G})$.
The {\em weight} of an [$f,g$]-factor ${\cal H}$ of ${\cal G}$ is the total weight of the edges in ${\cal H}$.

Let $C_1, C_2, \ldots, C_r$ be the $2$-cycles of $\mcC$. 
We construct an auxiliary edge-weighted undirected graph ${\cal G} = (V^+ \cup V^- \cup X, F_1 \cup F_2 \cup F_3)$ from $G_1 = (V(G), E_1)$ as follows
(see Figure~\ref{fig04} for an illustration): 
\begin{itemize}
\parskip=0pt
\item $V^+ = \{v^+~|~v\in V(G)\}$, $V^- = \{v^-~|~v\in V(G)\}$, and $X = \{x_i, y_i~|~1\le i\le r\}$. 
\item $F_1=\{\{u^+,v^-\}~|~(u,v)\in E_1\}$, $F_2 = \{ \{x_i, v^+\}, \{x_i, v^-\}~|~1\le i\le r, v \in V(C_i)\}$, and $F_3 = \{ \{x_i, y_i\}~|~1\le i\le r\}$. 
\item The weight of each edge in $F_1 \cup F_2$ is $0$ while the weight of each edge in $F_3$ is $1$. 
\item For each $v \in V(G) - \bigcup_{i=1}^r V(C_i)$, $f(v^+) = f(v^-) = 0$ and $g(v^+) = g(v^-) = 1$. 
\item For each $v \in \bigcup_{i=1}^r V(C_i)$, $f(v^+) = f(v^-) = g(v^+) = g(v^-) = 1$. 
\item For each $i\in\{1, 2, \ldots, r\}$, $f(x_i) = f(y_i) = 0$, $g(x_i) = 4$, and $g(y_i) = 1$. 
\end{itemize}

\begin{figure}[ht]
\centering
  \setlength{\unitlength}{1bp}%
  \begin{picture}(231.55, 159.08)(0,0)
  \put(0,0){\includegraphics{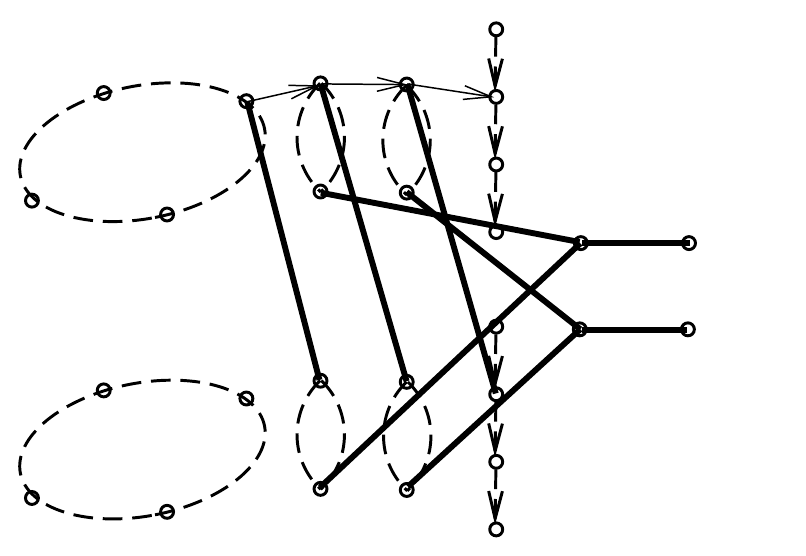}}
  \put(83.18,140.09){\fontsize{11.38}{13.66}\selectfont $v^+$}
  \put(53.52,120.31){\fontsize{11.38}{13.66}\selectfont $u^+$}
  \put(92.33,54.66){\fontsize{11.38}{13.66}\selectfont $v^-$}
  \put(53.52,34.68){\fontsize{11.38}{13.66}\selectfont $u^-$}
  \put(158.61,96.96){\fontsize{11.38}{13.66}\selectfont $x_1$}
  \put(193.65,96.96){\fontsize{11.38}{13.66}\selectfont $y_1$}
  \put(162.51,54.14){\fontsize{11.38}{13.66}\selectfont $x_2$}
  \put(197.54,54.14){\fontsize{11.38}{13.66}\selectfont $y_2$}
  \end{picture}%
\caption{An illustration of the construction of ${\cal G}$ from $G_1 = (V(G), E_1)$,
	where the edges in the $4$-component path-cycle cover $\mcC$ are shown dashed,
	the edges in a maximum weight path-cover $M$ of $G_1$ are shown thin solid directed,
	and the edges in the corresponding maximum weight $[f, g]$-factor $N$ of ${\cal G}$ are shown thick solid undirected.\label{fig04}}
\end{figure}
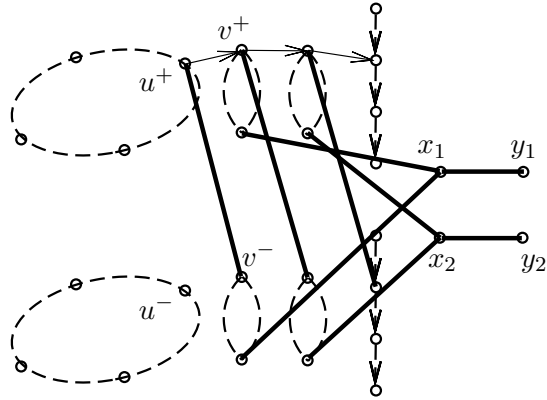

For each path-cycle cover $M$ of $G_1$, we can obtain an $[f,g]$-factor $N$ of ${\cal G}$ from $M$ as follows:
\begin{enumerate}
\parskip=0pt
\item Initially, $N = \{\{u^+,v^-\}~|~(u,v)\in E(M)\}$. 
\item For each $v \in \bigcup_{i=1}^r V(C_i)$ with $d^+_M(v) = 0$, add the edge $\{v^+, x_i\}$ to $N$. 
\item For each $v \in \bigcup_{i=1}^r V(C_i)$ with $d^-_M(v) = 0$, add the edge $\{v^-, x_i\}$ to $N$. 
\item For each $i \in \{1, 2, \ldots, r\}$ with $d_N(x_i) < 4$, add the edge $\{x_i, y_i\}$ to $N$.%
	\footnote{The degree of $x_i$ is either $3$ or $4$, corresponding to $2$ or $1$ edge of $M$ saturating the $2$-cycle $C_i$, respectively.}
\end{enumerate}
From the edge weight settings, the weight of $N$ is the same as that of $M$, which is equal to the number of $2$-cycles saturated by $M$.
Thus, the maximum weight of an $[f, g]$-factor of ${\cal G}$ is at least as large as the maximum weight of a path-cycle cover of $G_1$. 

Conversely, from each maximum weight $[f, g]$-factor $N$ of ${\cal G}$, we can obtain a path-cycle cover $M$ of $G_1$ by letting 
$E(M) = \{(u, v) \mid \{u^+, v^-\} \in E(N) \cap F_1\}$. 
We claim that the weight of $M$ is the same as that of $N$. 
To see this claim, observe that for each $i \in \{1, 2, \ldots, r\}$ such that at least one edge of $M$ is incident at a vertex $v$ of $C_i$, 
$N$ cannot contain both $\{v^+, x_i\}$ and $\{v^-, x_i\}$ due to the function $g$,
and in turn $\{x_i, y_i\}$ must be contained in $N$ because $N$ is a maximum weight $[f, g]$-factor of ${\cal G}$.
By the claim, the maximum weight of a path-cycle cover of $G_1$ is at least as large as the maximum weight of an $[f, g]$-factor of ${\cal G}$. 

We conclude that the maximum weight of a path-cycle cover of $G_1$ is the same as the maximum weight of an $[f, g]$-factor of ${\cal G}$.
This proves the lemma.
\end{proof}

An undirected graph ${\cal G}$ is a {\em star} if ${\cal G}$ is a connected graph with at least one edge and
all but at most one vertex of ${\cal G}$ have degree~1 in ${\cal G}$.
If a star ${\cal G}$ has a vertex of degree larger than~1, then this unique vertex is the {\em center} of ${\cal G}$;
otherwise, ${\cal G}$ is an edge and we choose an arbitrary vertex of ${\cal G}$ as the {\em center} of ${\cal G}$.
Each vertex of a star ${\cal G}$ other than its center is a {\em satellite} of ${\cal G}$.
A vertex $u$ is {\em isolated} in an undirected graph ${\cal G}$ if the degree of $u$ in ${\cal G}$ is~0.

Our algorithm then proceeds to perform the following four steps described in Figure~\ref{fig05},
in which another subgraph $G_2$ of $G$ and an undirected graph $G_3$ are constructed.

\begin{figure}[htb]
\begin{center}
\framebox{
\begin{minipage}{5.5in}
\begin{enumerate}
\parskip=0pt
\setcounter{enumi}{3} 
\item\label{step:M} 
	Compute a maximum-weight path-cycle cover $M$ in $G_1$ (via Lemma~\ref{lem:2match}). 
\item\label{step:min}
	While $\exists e \in E(M)$ such that its removal does not change the weight of $M$,
	\begin{itemize}
	\parskip=0pt
	\item[5.1]
		$M \leftarrow M - e$ (that is, delete $e$ from $M$).
	\end{itemize}
\item\label{step:G2}
	Construct a directed graph $G_2 = (V(G), E(\mcC) \cup E(M))$.\\ 
	({\em Comment:} For each pair of connected components of $\mcC$, there 
	is at most one edge between them in $G_2$ because of Step~\ref{step:min}.)
\item\label{step:G2'}
	Construct an undirected graph $G_3$, where the vertices of $G_3$ one-to-one 
	correspond to the connected components of $\mcC$ and two vertices 
	are adjacent in $G_3$ if and only if the corresponding connected 
	components of $\mcC$ are adjacent in $G_2$. 
\end{enumerate} 
\end{minipage}}
\end{center}
\caption{The continued description of {\sc Approx2}, the next four steps (to be continued).\label{fig05}}
\end{figure}

\begin{fact}
\label{fact01}
For each connected component $H$ of $G_3$,
	\begin{enumerate}
	\parskip=0pt
	\item $H$ is an isolated vertex or a star. 
	\item If $H$ contains at least three vertices, then every satellite of $H$ corresponds to a $2$-cycle of $\mcC$. 
	\item If $H$ contains two vertices, then at least one vertex of $H$ corresponds to a $2$-cycle of $\mcC$.
		({\em Comment:} In this case, we always choose a vertex corresponding to a $2$-cycle of $\mcC$ to be the satellite of $H$.)
	\end{enumerate}
\end{fact}
\begin{proof}
These facts are obvious because of the definition of $G_1$ and Step~\ref{step:min},
so that every $2$-cycle of $\mcC$ is incident with at most one edge of $M$ and
every other connected component of $\mcC$ can be adjacent to only $2$-cycles.
\end{proof}

An {\em isolated $2$-cycle} of $G_2$ is a $2$-cycle of $\mcC$ whose corresponding vertex in $G_3$ is isolated in $G_3$, otherwise a {\em leaf $2$-cycle}.
Let $\mcI$ be the set of isolated $2$-cycles in $G_2$.
Let $\mcQ^*$ be an optimal $k$-path partition of $G$.

\begin{lemma}\label{lem:isoCyc}
$|E(\mcQ^*)| \le \min\{|E(\mcC)| - |\mcI|, \frac{k-1}{k}\cdot(n - 2|\mcI|) + |\mcI|\}$.
\end{lemma}
\begin{proof}
Let $C_1$, \ldots, $C_h$ be those $2$-cycles of $\mcC$ such that for each $i \in \{1, \ldots, h\}$,
no edge of $E(\mcQ^*)$ is incident at exactly one vertex of $V(C_i)$
(in other words, due to the optimality of $\mcQ^*$, exactly one edge of $E(C_i)$ is a $2$-path in $\mcQ^*$).

Let $U_1 = \bigcup_{i=1}^h V(C_i)$ and $U_2 = V(G) - U_1$.
For convenience, let $C_0 = G[U_2]$.
Note that for each $e \in E(\mcQ^*)$, one of the subgraphs $C_0$, $C_1$, \ldots, $C_h$ contains both endpoints of $e$.
Therefore, $\mcQ^*$ can be partitioned into $h+1$ disjoint subgraphs $\mcQ^*_0$, \ldots, $\mcQ^*_h$ such that 
$\mcQ^*_i$ is a $k$-path partition (and hence a path-cycle cover) of $G[V(C_i)]$ for every $i \in \{0, \ldots, h\}$. 
Since $\mcC[U_2]$ must be a maximum path-cycle cover of $C_0$, $|E(\mcC[U_2])| \ge |E(\mcQ^*_0)|$.
Combining with the fact that, for every $i \in \{1, \ldots, h\}$, $|E(\mcQ^*_i)| = 1$,
we have
\begin{equation}
\label{eq01}
|E(\mcC)| = |E(\mcC[U_2])|+ \sum_{i=1}^h |E(C_i)| \ge |E(\mcQ^*_0)| + 2h = |E(\mcQ^*)| + h.
\end{equation}

Note that $(V(G), E(G_1) \cap E(\mcQ^*))$ is a path-cycle cover in $G_1$ of weight $r - h$, where $r$ is the total number of $2$-cycles in $\mcC$. 
One sees that $r - h \le r - |\mcI|$ because $M$ is a maximum-weight path-cycle cover in $G_1$ of weight $r - |\mcI|$.
That is,
\begin{equation}
\label{eq02}
|\mcI| \le h.
\end{equation}

By Eqs.~(\ref{eq01}) and (\ref{eq02}) we have $|E(\mcQ^*)| \le |E(\mcC)| - h \le |E(\mcC)| - |\mcI|$.
This establishes the first half of the lemma. 
On the other hand, since $\mcQ^*_0$ is a $k$-path partition of $C_0$, each path in $\mcQ^*_0$ can have at most $k-1$ edges and hence
$\mcQ^*_0$ can have at most $\frac{k-1}{k}\cdot |U_2| = \frac{k-1}{k}\cdot (n - 2h)$ edges.
Therefore, $E(\mcQ^*) = |E(\mcQ^*_0)| + h \le \frac{k-1}{k}\cdot (n - 2h) + h \le \frac{k-1}{k}\cdot (n - 2|\mcI|) + |\mcI|$,
where the last inequality holds by Eq.~(\ref{eq02}).
This establishes the second half of the lemma. 
\end{proof}

\begin{lemma}\label{lem:vertex}
Suppose the connected component $C$ of $\mcC$ corresponding to an isolated vertex of $G_3$ is not a $2$-cycle.
Then, $C$ can be transformed into a $k$-path partition $\mcP$ of $G[V(C)]$ such that $|E(\mcP)| \ge \frac{2}{3}\cdot |E(C)|$.
\end{lemma}
\begin{proof}
We distinguish two cases.
In the first case where $C$ is a path, we can transform $C$ into a $k$-path partition $\mcP$ of $G[V(C)]$ by
starting at one end of the path $C$ and deleting every $k$-th edge.
Clearly, $|E(\mcP)| \ge \frac{k-1}{k} \cdot |E(C)| \ge \frac 67 \cdot |E(C)|$ due to $k \ge 7$.

In the second case where $C$ is a cycle, we can transform $C$ into a $k$-path partition $\mcP$ of $G[V(C)]$ by
first deleting an arbitrary edge and then starting at one end of the resulting path and further deleting every $k$-th edge.
When $|E(C)| \ge 6$,
$|E(\mcP)| \ge \frac{k-1}{k} \cdot (|E(C)| - 1) \ge \frac 67 \cdot \frac 56 \cdot |E(C)| = \frac 57 \cdot |E(C)|$
due to $k \ge 7$;
otherwise, $|E(\mcP)| = |E(C)| - 1 \ge \frac{2}{3}\cdot|E(C)|$ since $|E(C)|\ge 3$. 
\end{proof}

\begin{lemma}\label{lem:path}
Suppose the connected component of $\mcC$ corresponding to the center of a star connected component of $G_3$ is a path.
Let $U = \bigcup_C V(C)$ and $F= \bigcup_C E(C)$, where $C$ ranges over all connected components of $\mcC$ corresponding to the vertices of the star.
Then, $G_2[U]$ can be transformed into a $k$-path partition $\mcP$ of $G[U]$ such that $|E(\mcP)| \ge \frac 23 \cdot |F|$. 
\end{lemma}
\begin{proof}
Let $P$ be the path connected component of $\mcC$ corresponding to the center of the star,
and let $C_1$, \ldots, $C_h$ be the $2$-cycles of $\mcC$ corresponding to all the satellites (see Fact~\ref{fact01}).
For each $i \in \{1, \ldots, h\}$, let $e_i$ denote the unique edge of $G_2[U]$ between a vertex $u_i$ of $C_i$ and a vertex $v_i$ of $P$.
Let $s$ ($t$, respectively) be the starting (ending, respectively) vertex of $P$;
and assume without loss of generality that $v_1, \ldots, v_h$ lie sequentially on $P$ (see Figure~\ref{fig06} for an illustration).
We note that either $e_i = (u_i, v_i)$ or $e_i = (v_i, u_i)$, and it is possible that $v_i = v_j$ for $i < j$.
Nevertheless, $v_i = v_j$ implies that one of $e_i$ and $e_j$ enters $v_i$ and the other leaves $v_i$, because $M$ is a path-cycle cover in $G_1$.
That is, at most two $2$-cycles can be adjacent to the same $v_i$.
For convenience, we say that a vertex $v$ of $P$ is {\em free} if $v \not \in\{v_1, \ldots, v_h\}$.

We prove the lemma by induction on $|V(P)|$.
In the base case, $|V(P)| = 1$ and thus $h \le 2$,
we can transform $G_2[U]$ into an $\ell$-path with $\ell \le 5$ by simply removing one appropriate edge from each $2$-cycle in $\{C_1, \ldots, C_h\}$.
This single $\ell$-path gives rise to a $k$-path partition $\mcP$ such that $|E(\mcP)| = |F|$.

Next, suppose $|V(P)| \ge 2$.
We distinguish four cases below (see Figure~\ref{fig06} for an illustration).

\begin{figure}[ht]
\centering
  \setlength{\unitlength}{1bp}%
  \begin{picture}(235.20, 125.65)(0,0)
  \put(0,0){\includegraphics{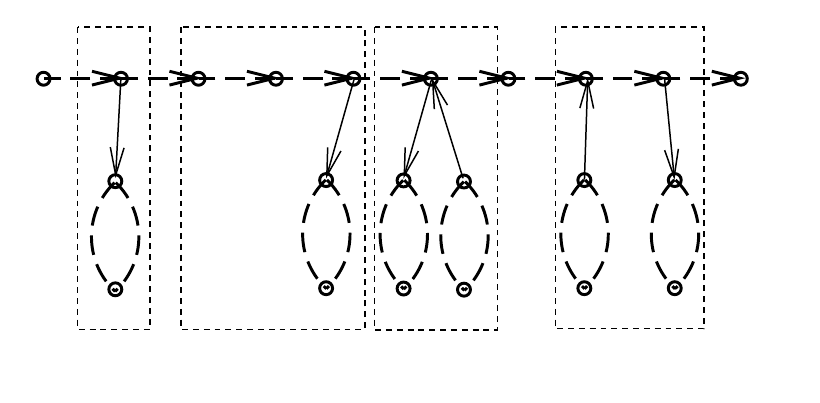}}
  \put(5.67,97.37){\fontsize{11.38}{13.66}\selectfont $s$}
  \put(215.04,96.72){\fontsize{11.38}{13.66}\selectfont $t$}
  \put(20.54,8.12){\fontsize{11.38}{13.66}\selectfont Case 4}
  \put(65.17,8.12){\fontsize{11.38}{13.66}\selectfont Case 2}
  \put(109.80,8.12){\fontsize{11.38}{13.66}\selectfont Case 1}
  \put(169.30,8.12){\fontsize{11.38}{13.66}\selectfont Case 3}
  \put(31.70,101.09){\fontsize{11.38}{13.66}\selectfont $v_1$}
  \put(98.64,101.09){\fontsize{11.38}{13.66}\selectfont $v_2$}
  \put(117.24,101.09){\fontsize{11.38}{13.66}\selectfont $v_3/v_4$}
  \put(165.58,101.09){\fontsize{11.38}{13.66}\selectfont $v_5$}
  \put(187.90,101.09){\fontsize{11.38}{13.66}\selectfont $v_6$}
  \end{picture}%
\caption{An illustration of the four distinct cases of how $2$-cycles of $\mcC$ are adjacent to a path component in $G_2$,
	where the dashed edges are in $E(\mcC)$ and the solid edges are in $E(M)$.\label{fig06}}
\end{figure}
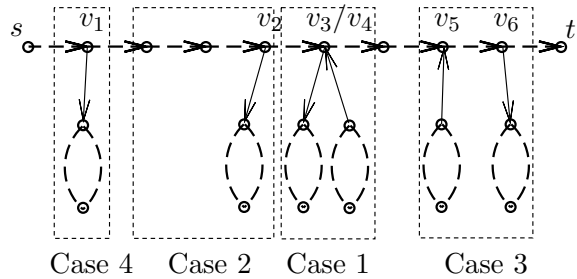

{\em Case 1:} There exists $i$ such that $1\le i < h$ and $v_i = v_{i+1}$.
In this case, we transform $G_2[U]$ into $\mcP$ as follows:
First, we remove one edge from $C_i$ and remove one edge from $C_{i+1}$ so that
$e_i$ and $e_{i+1}$ together with the remaining edges of $C_i$ and $C_{i+1}$ form a $5$-path denoted as $Q$.
Next remove all the edges of $P$ incident at $v_i$,
and let $Q$, ${\cal G}_1$ and ${\cal G}_2$ (${\cal G}_2$ could be empty) be the connected components of the resulting graph.
Lastly, for ${\cal G}_1$ and ${\cal G}_2$ we recursively transform each of them into a $k$-path partition $\mcP_j$ of $G[V({\cal G}_j)]$.
Obviously, $\mcP = \{Q\} \cup \mcP_1 \cup \mcP_2$ is a $k$-path partition of $G[U]$. 
By the inductive hypothesis, $|E(\mcP_j)| \ge \frac{2}{3}\cdot |E({\cal G}_j) \cap F|$ for $j = 1, 2$.
Therefore, $|E(\mcP)| = |E(Q)| + |E(\mcP_1)| + |E(\mcP_2)| \ge 4 + \frac{2}{3} (|E({\cal G}_1)\cap F| + |E({\cal G}_2)\cap F|)$. 
We also have $|F| \le 6 + |E({\cal G}_1)\cap F| + |E({\cal G}_2)\cap F|$.
It follows that $|E(\mcP)| \ge \frac{2}{3}\cdot |F|$.

{\em Case 2:} The path $P$ contains a sub-path $x$-$y$-$z$ such that
either $x$ and $y$ are free and $z = v_i \in \{v_1,\ldots,v_h\}$ with $e_i = (v_i, u_i)$
or $y$ and $z$ are free and $x = v_i \in \{v_1,\ldots,v_h\}$ with $e_i = (u_i, v_i)$.
We transform $G_2[U]$ into $\mcP$ as follows:
First, we remove one edge from $C_i$ so that $x$-$y$-$z$ and $e_i$ together with the remaining edge of $C_i$ form a $5$-path denoted as $Q$.
Next remove the edge of $P$ entering $x$ and remove the edge of $P$ leaving $z$, if any,
followed by applying the same inductive argument (except that ${\cal G}_1$ could also be empty) as in Case 1.

{\em Case 3:} The path $P$ contains a sub-path $P'$ from $v_i$ to $v_{i+1}$ for some $1 \le i < h$ such that
$e_i = (u_i, v_i)$ and $e_{i+1} = (v_{i+1}, u_{i+1})$.
We transform $G_2[U]$ into $\mcP$ as follows:
First, we remove one edge from $C_i$ and remove one edge from $C_{i+1}$ so that
$P'$, $e_i$, and $e_{i+1}$ together with the remaining edges of $C_i$ and $C_{i+1}$ form an $\ell$-path denoted as $Q$
(where $\ell = 6$ or $7$\footnote{This is why we need to assume $k \ge 7$ in this subsection.}). 
Next remove the edge of $P$ entering $v_i$ and remove the edge of $P$ leaving $v_{i+1}$, if any,
and let $Q$, ${\cal G}_1$ and ${\cal G}_2$ (${\cal G}_1$ and ${\cal G}_2$ could be empty) be the connected components of the resulting graph.
Lastly, for ${\cal G}_1$ and ${\cal G}_2$ we recursively transform each of them into a $k$-path partition $\mcP_j$ of $G[V({\cal G}_j)]$.
Obviously, $\mcP = \{Q\} \cup \mcP_1 \cup \mcP_2$ is a $k$-path partition of $G[U]$. 
By the inductive hypothesis, $|E(\mcP_j)| \ge \frac{2}{3}\cdot |E({\cal G}_j) \cap F|$ for $j = 1, 2$.
Therefore, $|E(\mcP)| = |E(Q)| + |E(\mcP_1)| + |E(\mcP_2)| \ge (\ell - 1) + \frac{2}{3} (|E({\cal G}_1)\cap F| + |E({\cal G}_2)\cap F|)$. 
We also have $|F| \le (\ell + 1) + |E({\cal G}_1)\cap F| + |E({\cal G}_2)\cap F|$.
It follows from $\ell = 6$ or $7$ that $|E(\mcP)| \ge \frac{2}{3}\cdot |F|$.

{\em Case 4:} The edge $e_1$ leaves $v_1$, or the edge $e_h$ enters $v_h$.
We assume the first scenario, that is, $e_1 = (v_1, u_1)$; 
the other scenario is symmetric.
By Case~2, the sub-path $P'$ of $P$ from $s$ to $v_1$ contains at most one edge. 
We transform $G_2[U]$ into $\mcP$ as follows:
First, we remove one edge from $C_1$ so that $P'$ and $e_1$ together with the remaining edge of $C_1$ form an $\ell$-path denoted as $Q$
(where $\ell = 3$ or $4$). 
Next remove the edge of $P$ leaving $v_1$, if any,
and let $Q$ and ${\cal G}_1$ (${\cal G}_1$ could be empty) be the connected components of the resulting graph.
Lastly, for ${\cal G}_1$ we recursively transform it into a $k$-path partition $\mcP_1$ of $G[V({\cal G}_1)]$.
Obviously, $\mcP = \{Q\} \cup \mcP_1$ is a $k$-path partition of $G[U]$. 
By the inductive hypothesis, $|E(\mcP_1)| \ge \frac{2}{3}\cdot |E({\cal G}_1) \cap F|$.
Therefore, $|E(\mcP)| = |E(Q)| + |E(\mcP_1)| \ge (\ell - 1) + \frac{2}{3} \cdot |E({\cal G}_1)\cap F|$. 
We also have $|F| \le \ell + |E({\cal G}_1)\cap F|$.
It follows from $\ell = 3$ or $4$ that $|E(\mcP)| \ge \frac{2}{3}\cdot |F|$.

We argue that there is no other case.
By Case~1, these $h$ vertices $v_1, \ldots, v_h$ are distinct.
Then by Case~4, $e_1$ enters $v_1$.
Next by Case~3, $e_2$ enters $v_2$ too;
and iteratively every $e_i \in \{e_1, \ldots, e_h\}$ enters $v_i$, which implies that Case~4 occurs.
This finishes the proof of the lemma.
\end{proof}

\begin{lemma}\label{lem:cycle}
Suppose the connected component of $\mcC$ corresponding to the center of a star connected component of $G_3$ is a cycle.
Let $U = \bigcup_C V(C)$ and $F= \bigcup_C E(C)$, where $C$ ranges over all connected components of $\mcC$ corresponding to the vertices of the star.
Then, $G_2[U]$ can be transformed into a $k$-path partition $\mcP$ of $G[U]$ such that $|E(\mcP)| \ge \frac 23 \cdot |F|$. 
\end{lemma}
\begin{proof}
We inherit the notations and definitions in the first paragraph of the proof of Lemma~\ref{lem:path},
to let $P$ denote the cycle connected component of $\mcC$ corresponding to the center of the star (see Fact~\ref{fact01}).
We prove the lemma by distinguishing the first three cases as in the inductive phase in the proof of Lemma~\ref{lem:path},
and replacing the inductive hypotheses by Lemma~\ref{lem:path}.
Indeed, for each case, the argument remains the same, except that there are exactly two connected components in the resulting graph,
one of which is $Q$ and the other ${\cal G}_1$ is non-empty.

{\em Case 4:} None of Cases~1--3 occurs.
By Case~1, these $h$ vertices $v_1, \ldots, v_h$ are distinct.
Then by Case~3, either every $e_i \in \{e_1, \ldots, e_h\}$ leaves $v_i$ or every $e_i \in \{e_1, \ldots, e_h\}$ enters $v_i$.
We assume the former scenario;
the latter is symmetric.
We transform $G_2[U]$ into $\mcP$ as follows:
First, we remove the edge of $C_i$ entering $u_i$;
then remove the edge of $P$ leaving $v_i$, for every $i \in \{1, \ldots, h\}$.
Let $Q_1$, \ldots, $Q_h$ be the connected components of the resulting graph, each is a path of order $3$ or $4$ by Case~2.
Therefore, $\mcP = \{Q_1, \ldots, Q_\ell\}$ is a $k$-path partition of $G[U]$.
Moreover, $|F| = |E(P)| + 2h$ and $|E(\mcP)| = |F| - h = |E(P)| + h$.
It follows from $|E(P)| \ge h$ that $|E(\mcP)| \ge \frac{2}{3} \cdot |F|$.  
This finishes the proof of the lemma.
\end{proof}

Now we are ready to state the final four steps of our algorithm {\sc Approx2} in Figure~\ref{fig07},
and summarize the result in the following theorem.

\begin{figure}[htb]
\begin{center}
\framebox{
\begin{minipage}{5.5in}
\begin{enumerate}
\parskip=0pt
\setcounter{enumi}{7} 
\item\label{step:iso} 
	For each isolated vertex of $G_3$ corresponding to a $2$-cycle $C$ in $\mcC$,
	transform $C$ into a $k$-path partition of $G[V(C)]$ by deleting an edge of $C$. 
\item\label{step:vertex} 
	For each isolated vertex of $G_3$ not corresponding to a $2$-cycle in $\mcC$,
	use the algorithm implied by 	Lemma~\ref{lem:vertex} to transform $G_2[U]$ into a $k$-path partition of $G[U]$,
	where $U$ is defined as in Lemma~\ref{lem:vertex}. 
\item\label{step:star} 
	For each star connected component of $G_3$ whose center corresponds to a path/cycle of $\mcC$,
	use the algorithm implied by 	Lemma~\ref{lem:path}/\ref{lem:cycle} to transform $G_2[U]$ into a $k$-path partition of $G[U]$,
	where $U$ is defined as in Lemma~\ref{lem:path}/\ref{lem:cycle}.
\item 
	Return the union $\mcQ$ of the $k$-path partitions obtained in Steps~\ref{step:iso}--\ref{step:star}.	
\end{enumerate} 
\end{minipage}}
\end{center}
\caption{The continued description of {\sc Approx2}.\label{fig07}}
\end{figure}

\begin{theorem}
\label{thm02}
The algorithm {\sc Approx2} is an $O(nm \log n)$-time $(k+2)/3$-approximation for the $k$PP problem, where $k \ge 7$.
\end{theorem}
\begin{proof}
The time complexity of our algorithm {\sc Approx2} is dominated by the computation of
the maximum path-cycle cover $\mcC$ in $O(n m \log n)$ time~\cite{Gab83},
and the maximum weight path-cycle cover $M$ which is done in $O(n m \log n)$ time by Lemma~\ref{lem:2match}.

To analyze the approximation ratio, let $\mcQ$ be the $k$-path partition returned by our algorithm.
We have
\begin{itemize}
\parskip=0pt
\item
	$|E(\mcC)| \ge |E(\mcQ^*)|$, obviously;
\item
	$|E({\cal Q})| \ge |\mcI| + \frac{2}{3}\left(|E(\mcC)| - 2|\mcI|\right)
		= \frac{2}{3}|E(\mcC)| - \frac{1}{3}|\mcI| \ge \frac{2}{3}|E(\mcQ^*)| - \frac{1}{3}|\mcI|$ by Lemmas~\ref{lem:vertex}--\ref{lem:cycle}; 
\item
	$|E(\mcQ^*)| \le \frac{k-1}{k}\cdot(n - 2|\mcI|) + |\mcI| = \frac{k-1}{k}n - \frac{k-2}{k}|\mcI|$ by Lemma~\ref{lem:isoCyc};
\item
	$|\mcQ| = n - |E({\cal Q})|$ and $|\mcQ^*| = n - |E(\mcQ^*)|$.
\end{itemize}
The approximation ratio is
$\frac{|\mcQ|}{|\mcQ^*|} = \frac{n-|E({\cal Q})|}{n-|E(\mcQ^*)|} \le \frac{n - \frac{2}{3}|E(\mcQ^*)| + \frac{1}{3}|\mcI|} {n-|E(\mcQ^*)|}$.
One can verify that the last fraction is an increasing function in $|E(\mcQ^*)|$.
Therefore, 
\[
	\frac{|\mcQ|}{|\mcQ^*|}
\le \frac{n - \frac{2}{3}\left(\frac{k-1}{k}n - \frac{k-2}{k}|\mcI|\right) + \frac{1}{3}|\mcI|} {n-\left(\frac{k-1}{k}n - \frac{k-2}{k}|\mcI|\right)}
= \frac{(k+2)n + (3k-4)|\mcI|} {3n + (3k-6)|\mcI|}
\le \max\left\{\frac{k+2}{3}, \frac{3k-4}{3k-6}\right\}
= \frac{k+2}{3},
\]
where the last equality is due to $k \ge 7$.
That is, our algorithm {\sc Approx2} is a $\frac{k+2}{3}$-approximation for the $k$PP problem where $k \ge 7$.
\end{proof}

\subsection{A $13/9$-approximation for $3$PP}\label{sec23}
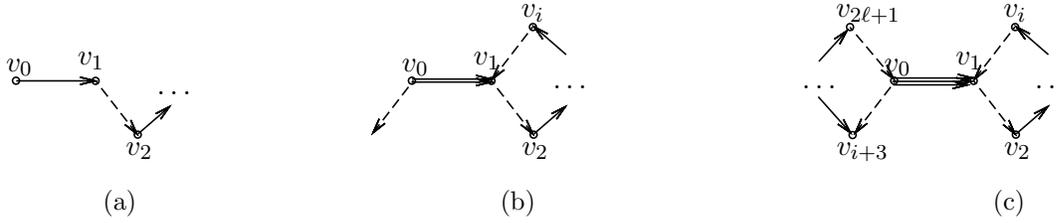
\begin{figure}[ht]
\centering
\begin{subfigure}{0.20\textwidth}
  \setlength{\unitlength}{0.7bp}%
  \begin{picture}(129.16, 96.72)(0,0)
  \put(0,0){\includegraphics[scale=0.7]{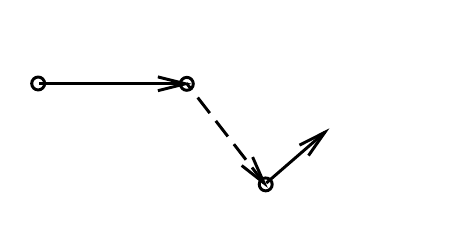}}
  \put(5.67,53.14){\fontsize{11.38}{13.66}\selectfont $v_0$}
  \put(44.21,56.46){\fontsize{11.38}{13.66}\selectfont $v_1$}
  \put(70.20,8.12){\fontsize{11.38}{13.66}\selectfont $v_2$}
  \put(86.95,40.40){\fontsize{11.38}{13.66}\selectfont $\ldots$}
  \end{picture}%
\caption{\label{fig08a}}
\end{subfigure}
\hspace{0.08\textwidth}
\begin{subfigure}{0.25\textwidth}
  \setlength{\unitlength}{0.7bp}%
  \begin{picture}(144.70, 96.72)(0,0)
  \put(0,0){\includegraphics[scale=0.7]{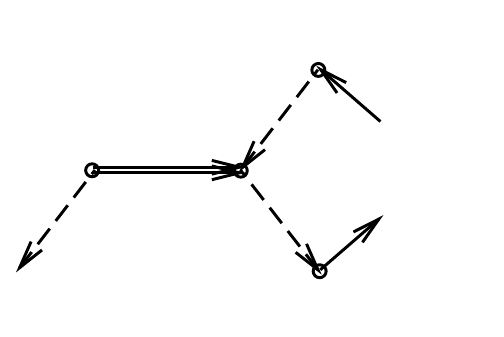}}
  \put(21.21,53.14){\fontsize{11.38}{13.66}\selectfont $v_0$}
  \put(59.75,56.46){\fontsize{11.38}{13.66}\selectfont $v_1$}
  \put(85.74,8.12){\fontsize{11.38}{13.66}\selectfont $v_2$}
  \put(102.48,43.61){\fontsize{11.38}{13.66}\selectfont $\ldots$}
  \put(85.44,82.16){\fontsize{11.38}{13.66}\selectfont $v_i$}
  \end{picture}%
\caption{\label{fig08b}}
\end{subfigure}
\hspace{0.08\textwidth}
\begin{subfigure}{0.35\textwidth}
  \setlength{\unitlength}{0.7bp}%
  \begin{picture}(173.30, 96.72)(0,0)
  \put(0,0){\includegraphics[scale=0.7]{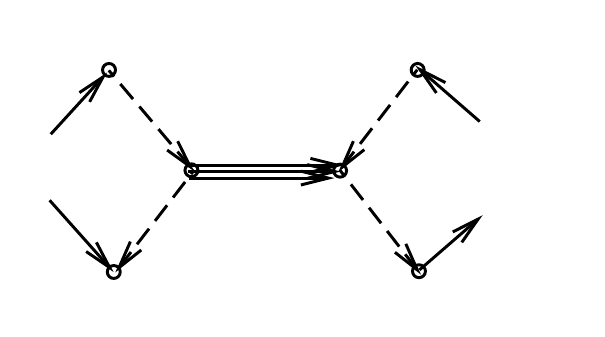}}
  \put(49.81,53.14){\fontsize{11.38}{13.66}\selectfont $v_0$}
  \put(88.35,56.46){\fontsize{11.38}{13.66}\selectfont $v_1$}
  \put(114.34,8.12){\fontsize{11.38}{13.66}\selectfont $v_2$}
  \put(131.08,43.61){\fontsize{11.38}{13.66}\selectfont $\ldots$}
  \put(114.04,82.16){\fontsize{11.38}{13.66}\selectfont $v_i$}
  \put(24.13,8.12){\fontsize{11.38}{13.66}\selectfont $v_{i+3}$}
  \put(5.67,43.61){\fontsize{11.38}{13.66}\selectfont $\ldots$}
  \put(23.83,82.16){\fontsize{11.38}{13.66}\selectfont $v_{2\ell+1}$}
  \end{picture}%
\caption{\label{fig08c}}
\end{subfigure}
\caption{Local configurations of an augmenting path:
	(a) The first matched edge $(v_0, v_1)$ and the first free edge $(v_1, v_2)$ form a $3$-path $v_0$-$v_1$-$v_2$ in the graph $G$;
	symmetrically, if the last matched edge is not included twice, then it and the last free edge form a $3$-path.
	(b) The first matched edge $(v_0, v_1)$ is included twice;
	the first free edge $(v_1, v_2)$ and its adjacent free edge $(v_i, v_1)$ on the augmenting path form a $3$-path $v_i$-$v_1$-$v_2$.
	(c) The first and the last matched edges are the same edge $(v_0, v_1)$;
	the first free edge $(v_1, v_2)$ and its adjacent free edge $(v_i, v_1)$ on the augmenting path form a $3$-path $v_i$-$v_1$-$v_2$,
	and the last free edge $(v_{2\ell+1}, v_0)$ and its adjacent free edge $(v_0, v_{i+3})$ on the augmenting path form
	a $3$-path $v_{2\ell+1}$-$v_0$-$v_{i+3}$.\label{fig08}}
\end{figure}

In this section, we present another kind of alternating and augmenting paths to reduce the number of $2$-paths in a $3$-path partition $\mcQ$.
This time, the edges on the $2$-paths of $\mcQ$ are {\em matched} edges and the edges outside of $\mcQ$ are {\em free edges}.
(The edges on the $3$-paths of $\mcQ$ are {\em irrelevant}.)
An augmenting path starts with a matched edge then a free edge, which form a $3$-path,
alternating matched and free edges, and lastly ends with a matched edge.
The intention is to convert three $2$-paths of $\mcQ$ into two $3$-paths,
and thus the augmenting path should contain at least three distinct matched edges.
Formally, the following constraints must be satisfied:
\begin{itemize}
\parskip=-2pt
\item[1)]
	An augmenting path starts with a matched edge, alternating free and matched edges,
	ends with a matched edge, and contains at least three distinct matched edges;
\item[2)]
	only the first and the last matched edges can be included twice (Figure~\ref{fig08b}),
	and if they are the same edge, then this edge is included exactly three times (Figure~\ref{fig08c});
\item[3)]
	the first matched edge and the first free edge form a $3$-path in the graph $G$ (Figure~\ref{fig08a});
\item[4)]
	if the first matched edge is included twice,
	then the first free edge and its adjacent free edge on the augmenting path form a $3$-path (Figure~\ref{fig08b});
\item[5)]
	if the last matched edge is not included twice, then it and the last free edge form a $3$-path (Figure~\ref{fig08a});
\item[6)]
	if the last matched edge is included twice,
	then the last free edge and its adjacent free edge on the augmenting path form a $3$-path (Figure~\ref{fig08b}).
\end{itemize}

Upon an augmenting path, an analogous symmetric difference adds its free edges to while removes its internal matched edges from $\mcQ$.
This way, the collection of $\ell$ $2$-paths of $\mcQ$ is transferred into $(\ell-3)$ $2$-paths and two $3$-paths on the same set of vertices,
here $\ell$ denotes the number of distinct matched edges on the augmenting path.
We point out that, if the first/last matched edge is included twice,
then it is removed during the symmetric difference since one copy is internal on the augmenting path.
One sees that the net effect is to transfer three $2$-paths into two $3$-paths, thus reducing the size of the path partition by $1$.
Also, during such processes, no singletons or existing $3$-paths of $\mcQ$ are touched.
We initialize the $3$-path partition $\mcQ$ to be the solution produced by the algorithm {\sc Approx1}.

To find an augmenting path, we define below the alternating paths,
each of which starts with a matched edge and then a free edge such that these two edges form a $3$-path in the graph $G$
(i.e., satisfying Constraint \#3).
Consider w.l.o.g. a matched edge $(v_0, v_1)$ followed by a free edge $(v_1, v_2)$, see Figure~\ref{fig08} for an illustration.
If the vertex $v_2$ is not incident with a matched edge, then the alternating path is not extendable;
or otherwise the unique matched edge incident at $v_2$ extends the alternating path.
Note that there is no direction requirement on the second matched edge and we w.l.o.g. assume it is $(v_2, v_3)$.
Next, similarly, if the vertex $v_3$ is not incident with any free edge, then the alternating path is not extendable;
or otherwise a free edge incident at $v_3$ extends the alternating path and the extending process goes on.
We remark that $v_2$ can collide into $v_0$, and if so (i.e., $(v_1, v_0)$ is a free edge),
then the edge $(v_0, v_1)$ is included the second time into the alternating path and
the two free edges incident at the vertex $v_1$ (either added already or to be added next) must form a $3$-path in the graph $G$,
that is, Constraint \#4 must be satisfied for the extending process to go on.
Indeed, during the extending process,
whenever a matched edge $e$ is included, Constraint \#2 is checked and there are three possible cases:
\begin{description}
\parskip=-2pt
\item[Case 1.]
	$e$ appears the first time on the alternating path. Then Constraints \#1, \#5 are checked:
\item[\hspace{0.2in} 1.1.]
	If both satisfied, then an augmenting path is achieved;
\item[\hspace{0.2in} 1.2.]
	otherwise the extending process goes on.
\item[Case 2.]
	$e$ appears the second time on the alternating path.
\item[\hspace{0.2in} 2.1.]
	If $e$ is the same as the first matched edge and Constraint \#4 is or can be satisfied,
	then the extending process goes on (and use a free edge to satisfy Constraint \#4, if necessary);
\item[\hspace{0.2in} 2.2.]
	if $e$ is not the same as the first matched edge and Constraint \#6 is satisfied, then an augmenting path is achieved;
\item[\hspace{0.2in} 2.3.]
	otherwise the extending process terminates.
\item[Case 3.]
	$e$ appears the third time on the alternating path. Then Constraints \#4, \#6 are checked:
\item[\hspace{0.2in} 3.1.]
	If both satisfied, then an augmenting path is achieved;
\item[\hspace{0.2in} 3.2.]
	otherwise the extending process terminates.
\end{description}

In the above, by ``the extending process goes on'' we mean to use a free edge incident at the last vertex to extend the alternating path,
with an additional consideration in Case 2.1 where such an edge might have to satisfy Constraint \#4.
And then the matched edge incident at the last vertex extends the alternating path.
At the non-existence of such a free edge or such a matched edge, the alternating path is not extendable.
By ``the extending process terminates'', we mean that the alternating path does not lead to an augmenting path and thus the process is early terminated.

\begin{lemma}
\label{lemma06}
Given a directed graph $G = (V, E)$ and a $3$-path partition $\mcQ$ with the minimum number of singletons,
determining whether or not there exists an augmenting path with respect to $\mcQ$, and if so finding one such path, can be done in $O(nm)$ time.
\end{lemma}
\begin{proof}
A single BFS traversal is sufficient to explore all alternating paths starting with a specific matched edge.
Note that there are at most $O(n)$ matched edges.
\end{proof}

\begin{lemma}
\label{lemma07}
Given a directed graph $G = (V, E)$ and a $3$-path partition $\mcQ$ with the minimum number of singletons,
if there exists an augmenting path with respect to $\mcQ$,
then $\mcQ$ can be transferred into another $3$-path partition with the same number of singletons,
three less $2$-paths, and two more $3$-paths in $O(n)$ time.
\end{lemma}
\begin{proof}
Let $P$ denote the augmenting path with respect to $\mcQ$, and $e$ denote its starting matched edge and $e'$ denote its ending matched edge.
Assume $P$ contains in total $\ell$ matched edges.
If all these $\ell$ edges are distinct,
then the symmetric difference adds the $\ell-1$ free edges to while removes the $\ell-2$ internal matched edges from $\mcQ$,
resulting in two new $3$-paths and $\ell-3$ new $2$-paths.
That is, the achieved $3$-path partition contains two more $3$-paths and three less $2$-paths.

If $e \ne e'$ and exactly one of them is included twice,
then $P$ contains $\ell-1$ distinct matched edges and
the symmetric difference adds the $\ell-1$ free edges to while removes the $\ell-2$ distinct matched edges from $\mcQ$,
resulting in two new $3$-paths and $\ell-4$ new $2$-paths.
That is, the achieved $3$-path partition contains two more $3$-paths and three less $2$-paths.

If $e \ne e'$ and both of them are included twice, or if $e = e'$,
then $P$ contains $\ell-2$ distinct matched edges and
the symmetric difference adds the $\ell-1$ free edges to while removes all its matched edges from $\mcQ$,
resulting in two new $3$-paths and $\ell-5$ new $2$-paths.
That is, the achieved $3$-path partition contains two more $3$-paths and three less $2$-paths.

One clearly sees that the transferring process takes $O(n)$ time by walking through the augmenting path $P$ once.
\end{proof}

\begin{figure}[htb]
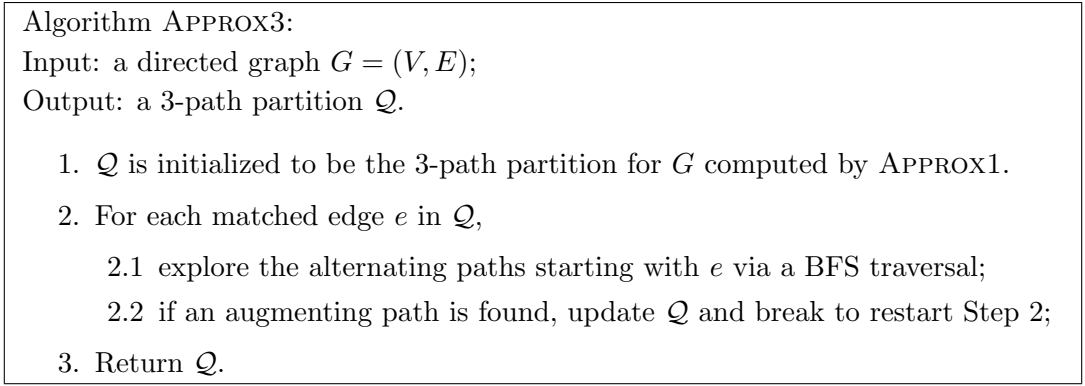

\begin{center}
\framebox{
\begin{minipage}{5.5in}
Algorithm {\sc Approx3}:\\
Input: a directed graph $G = (V, E)$;\\
Output: a $3$-path partition $\mcQ$.
\begin{itemize}
\parskip=0pt
\item[1.]
	$\mcQ$ is initialized to be the $3$-path partition for $G$ computed by {\sc Approx1}.
\item[2.]
	For each matched edge $e$ in $\mcQ$,
	\begin{itemize}
	\parskip=0pt
	\item[2.1]
		explore the alternating paths starting with $e$ via a BFS traversal;
	\item[2.2]
		if an augmenting path is found, update $\mcQ$ and break to restart Step 2;
	\end{itemize}
\item[3.]
	Return $\mcQ$.
\end{itemize}
\end{minipage}}
\end{center}
\caption{A high level description of the algorithm {\sc Approx3}.\label{fig09}}
\end{figure}

Our algorithm {\sc Approx3} starts with the $3$-path partition $\mcQ$ returned by the algorithm {\sc Approx1} for the input directed graph $G = (V, E)$.
Recall that $\mcQ$ contains the minimum number of singletons.
The same as {\sc Approx1}, the algorithm {\sc Approx3} is iterative too,
and in each iteration it tries to find an augmenting path starting with a matched edge with respect to the current $3$-path partition $\mcQ$,
and uses it to update $\mcQ$ to have the same number of singletons, three less $2$-paths, and two more $3$-paths.
To this purpose, the algorithm explores all the alternating paths starting with a matched edge,
and the iteration terminates at any time when an augmenting path is found, followed by updating the $3$-path partition.
If in an iteration no augmenting path is found for any matched edge,
then the algorithm {\sc Approx3} terminates and returns the achieved $3$-path partition as the solution.
From Lemmas~\ref{lemma06} and \ref{lemma07}, the overall running time is $O(n^2m)$.
A high level description of the algorithm is depicted in Figure~\ref{fig09}.

\begin{theorem}
\label{thm03}
The algorithm {\sc Approx3} is an $O(n^2m)$-time $13/9$-approximation for the $3$PP problem on directed graphs.
\end{theorem}
\begin{proof}
The running time of the algorithm is obvious from Theorem~\ref{thm01} and Lemmas~\ref{lemma06} and \ref{lemma07}.

Given a directed graph $G = (V, E)$,
note that the algorithm {\sc Approx3} starts with the $3$-path partition $\mcQ$ of $G$ computed by the algorithm {\sc Approx1}
and we have proved inside Theorem~\ref{thm01} that $\mcQ$ contains the minimum number of singletons among all $3$-path partitions.
Later improvements via augmenting paths do not touch any existing singleton or generate any new singleton,
and therefore the final $3$-path partition, still denoted as $\mcQ$, contains the minimum number of singletons too.

Let $\mcQ^*$ denote an optimal $3$-path partition that minimizes the number of paths.
Also, let $\mcQ^*_i$ ($\mcQ_i$, respectively) denote the sub-collection of all the $i$-paths of $\mcQ^*$ ($\mcQ$, respectively), for $i = 1, 2, 3$.
It follows that
\[
|\mcQ_1| + 2 |\mcQ_2| + 3 |\mcQ_3| = n = |\mcQ^*_1| + 2 |\mcQ^*_2| + 3 |\mcQ^*_3| \mbox{ and } |\mcQ_1| \le |\mcQ^*_1|.
\]
Below we do a counting to prove that $|\mcQ_2| \le |\mcQ^*_1| + 2 |\mcQ^*_2| + 4 |\mcQ^*_3|/3$.
Adding these three (in-) equalities together gives us
\[
3|\mcQ_1| + 3 |\mcQ_2| + 3 |\mcQ_3| \le 3 |\mcQ^*_1| + 4 |\mcQ^*_2| + 13 |\mcQ^*_3|/3,
\]
that is, $|\mcQ| \le 13 |\mcQ^*|/9$, and thus the theorem is proved.

Firstly, a singleton of $\mcQ^*_1$ is incident with at most one edge of $\mcQ_2$.
Equivalently speaking, the number of edges of $\mcQ_2$ that are incident at the singletons of $\mcQ^*_1$ is at most $|\mcQ^*_1|$.
Similarly, the number of edges of $\mcQ_2$ that are incident at the $2$-paths of $\mcQ^*_2$ is at most $2 |\mcQ^*_2|$.
Each of the other edges of $\mcQ_2$ has both its vertices on the $3$-paths of $\mcQ^*_3$,
and they are the matched edges with respect to the final $3$-path partition $\mcQ$.
Every edge of $\mcQ^*_3$ is free unless it is a matched edge.
We want to count the number of matched edges,
using the fact that there is no augmenting path in the subgraph of $G$ induced by the above defined matched and free edges, i.e., $E(\mcQ_2) \cup E(\mcQ^*_3)$.

In the following, for each matched edge incident at the mid-vertex of a $3$-path of $\mcQ^*_3$, we define an alternating path starting with it.

Assume there is a $3$-path $P_1 \in \mcQ^*_3$ of which the mid-vertex is incident with a matched edge denoted as $e^2_1$
(the other two matched edges incident at the vertices of $P_1$, if any, are denoted as $e^1_1$ and $e^3_1$, respectively;
see Figure~\ref{fig10} for an illustration).
The edge $e^2_1$ and exactly one of the two free edges on $P_1$, denoted as $f^1_1$, form a $3$-path in the graph $G$,
and consequently they can start an alternating path denoted as $\mcP$.
Either $\mcP$ is not extendable, 
or assume w.l.o.g. the matched edge $e^1_1$ extends $\mcP$ and $e^1_1$ is incident at a vertex of another $3$-path $P_2 \in \mcQ^*_3$ distinct from $P_1$.
One sees that if $e^1_1$ is incident at an end-vertex of $P_2$,
then exactly one free edge of $P_2$, denoted as $f^1_2$, extends $\mcP$ (Figure~\ref{fig10a});
if $e^1_1$ is incident at the mid-vertex of $P_2$,
then below we will choose exactly one of the two free edges of $P_2$ to extend $\mcP$ (Figure~\ref{fig10b}).

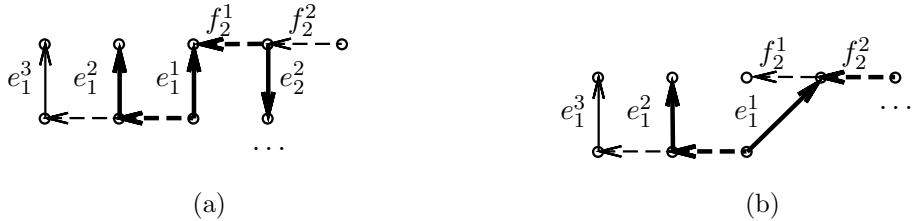
\begin{figure}[ht]
\centering
\begin{subfigure}{0.35\textwidth}
  \setlength{\unitlength}{1bp}%
  \begin{picture}(154.57, 68.77)(0,0)
  \put(0,0){\includegraphics{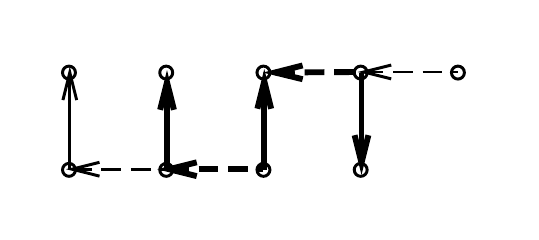}}
  \put(30.56,32.43){\fontsize{11.38}{13.66}\selectfont $e^2_1$}
  \put(97.99,8.12){\fontsize{11.38}{13.66}\selectfont $\ldots$}
  \put(61.68,32.43){\fontsize{11.38}{13.66}\selectfont $e^1_1$}
  \put(5.67,32.43){\fontsize{11.38}{13.66}\selectfont $e^3_1$}
  \put(80.35,54.21){\fontsize{11.38}{13.66}\selectfont $f^1_2$}
  \put(111.46,54.21){\fontsize{11.38}{13.66}\selectfont $f^2_2$}
  \put(108.35,32.43){\fontsize{11.38}{13.66}\selectfont $e^2_2$}
  \end{picture}%
\caption{\label{fig10a}}
\end{subfigure}
\hspace{0.08\textwidth}
\begin{subfigure}{0.35\textwidth}
  \setlength{\unitlength}{1bp}%
  \begin{picture}(167.97, 68.77)(0,0)
  \put(0,0){\includegraphics{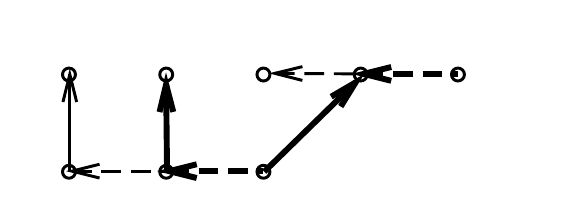}}
  \put(30.56,19.88){\fontsize{11.38}{13.66}\selectfont $e^2_1$}
  \put(125.76,23.80){\fontsize{11.38}{13.66}\selectfont $\ldots$}
  \put(71.13,19.88){\fontsize{11.38}{13.66}\selectfont $e^1_1$}
  \put(5.67,19.88){\fontsize{11.38}{13.66}\selectfont $e^3_1$}
  \put(80.35,41.66){\fontsize{11.38}{13.66}\selectfont $f^1_2$}
  \put(111.46,41.66){\fontsize{11.38}{13.66}\selectfont $f^2_2$}
  \end{picture}%
\caption{\label{fig10b}}
\end{subfigure}
\caption{Defining alternating paths: $\mcP$ starts with the matched edge incident at the mid-vertex of a $3$-path $P_1 \in \mcQ^*_3$.
	(a) The second matched edge $e^1_1$ of $\mcP$ is incident at an end-vertex of another $3$-path $P_2 \in \mcQ^*_3$,
	then the free edge on $P_2$ adjacent to $e^1_1$, which is $f^1_2$, extends $\mcP$.
	(b) The second matched edge $e^1_1$ of $\mcP$ is incident at the mid-vertex of another $3$-path $P_2 \in \mcQ^*_3$,
	then the free edge on $P_2$ that does not form together with $e^1_1$ a $3$-path in the graph $G$, which is $f^2_2$, extends $\mcP$.\label{fig10}}
\end{figure}

In the former case, either $\mcP$ is not extendable, 
or there is a matched edge incident at the mid-vertex of $P_2$, denoted as $e^2_2$, which extends $\mcP$ (Figure~\ref{fig10a}).
Denote the other free edge on $P_2$ as $f^2_2$.
We claim that there is no alternating path $\mcP'$ which includes $f^2_2$ and then includes $e^2_2$.
The reason is that one of $f^1_2$ and $f^2_2$ forms together with $e^2_2$ into a $3$-path in the graph $G$.
If $e^2_2$ is included the first time in $\mcP$ and the first time in $\mcP'$,
then one of $\mcP$ and $\mcP'$ is an augmenting path, which is a contradiction.
If $e^2_2$ is included the second time in at least one of $\mcP$ and $\mcP'$, then the concatenation of $\mcP$ and $\mcP'$ is an augmenting path,
which is a contradiction too.

In the latter case, $e^1_1$ and one of the two free edges on $P_2$, denoted as $f^1_2$, form a $3$-path in the graph $G$ (Figure~\ref{fig10b}).
Then the other free edge on $P_2$, denoted as $f^2_2$, extends $\mcP$.

Note that in both cases, the matched edge incident at the mid-vertex of $P_2$ and the other free edge on $P_2$ not included in $\mcP$
start another alternating path.
Nevertheless, in summary, we have ensured that no two alternating paths share any common free edge.
Due to the non-existence of an augmenting path,
every alternating path ends with a free edge on some $3$-path of $\mcQ^*_3$, such that this free edge is adjacent to only one matched edge
(otherwise the alternating path can be further extended).
That is, at most two vertices of this $3$-path of $\mcQ^*_3$ are incident with a matched edge each.
Let $\mcQ^*_{3,3}$ denote the sub-collection of $\mcQ^*_3$, each vertex on a $3$-path of which is incident with a matched edge.
Since each $3$-path of $\mcQ^*_{3,3}$ starts an alternating path but ends no alternating path, and
each $3$-path of $\mcQ^*_3 \setminus \mcQ^*_{3,3}$ ends at most two alternating paths,
we have
\[
|\mcQ^*_{3,3}| \le 2 (|\mcQ^*_3| - |\mcQ^*_{3,3}|),
\]
or equivalently
\[
|\mcQ^*_{3,3}| \le 2|\mcQ^*_3|/3.
\]
Therefore, the total number of matched edges incident at the $3$-paths of $\mcQ^*_3$ is at most
$(3 \times 2|\mcQ^*_3|/3 + 2 \times |\mcQ^*_3|/3) / 2 = 4|\mcQ^*_3|/3$.

Using the three estimates on the matched edges together,
we have proved that the total number of matched edges is at most $|\mcQ^*_1| + 2 |\mcQ^*_2| + 4 |\mcQ^*_3|/3$.
This finishes the proof of the theorem.
\end{proof}

\subsubsection{A tight instance for {\sc Approx3}}
The tight instance for the $13/9$-approximation algorithm for $3$PP on undirected graphs in \cite{CGL19a}
can be modified to show the tightness of {\sc Approx3}.

\begin{figure}[ht]
\centering
  \setlength{\unitlength}{1bp}%
  \begin{picture}(262.78, 92.86)(0,0)
  \put(0,0){\includegraphics{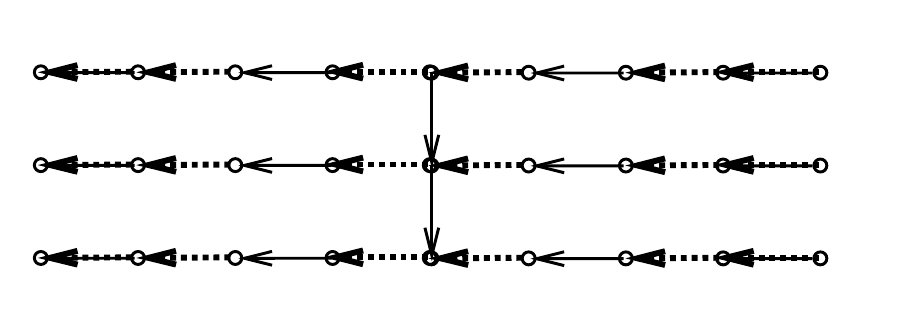}}
  \put(226.25,78.30){\fontsize{11.38}{13.66}\selectfont $u_0$}
  \put(199.51,78.30){\fontsize{11.38}{13.66}\selectfont $u_1$}
  \put(172.77,78.30){\fontsize{11.38}{13.66}\selectfont $u_2$}
  \put(146.04,78.30){\fontsize{11.38}{13.66}\selectfont $u_3$}
  \put(119.30,78.30){\fontsize{11.38}{13.66}\selectfont $u_4$}
  \put(89.22,78.30){\fontsize{11.38}{13.66}\selectfont $u_5$}
  \put(62.48,78.30){\fontsize{11.38}{13.66}\selectfont $u_6$}
  \put(35.75,78.30){\fontsize{11.38}{13.66}\selectfont $u_7$}
  \put(5.67,78.30){\fontsize{11.38}{13.66}\selectfont $u_8$}
  \put(226.25,51.57){\fontsize{11.38}{13.66}\selectfont $v_0$}
  \put(199.51,51.57){\fontsize{11.38}{13.66}\selectfont $v_1$}
  \put(172.77,51.57){\fontsize{11.38}{13.66}\selectfont $v_2$}
  \put(146.04,51.57){\fontsize{11.38}{13.66}\selectfont $v_3$}
  \put(111.00,51.57){\fontsize{11.38}{13.66}\selectfont $v_4$}
  \put(89.22,51.57){\fontsize{11.38}{13.66}\selectfont $v_5$}
  \put(62.48,51.57){\fontsize{11.38}{13.66}\selectfont $v_6$}
  \put(35.75,51.57){\fontsize{11.38}{13.66}\selectfont $v_7$}
  \put(5.67,51.57){\fontsize{11.38}{13.66}\selectfont $v_8$}
  \put(226.25,8.12){\fontsize{11.38}{13.66}\selectfont $w_0$}
  \put(199.51,8.12){\fontsize{11.38}{13.66}\selectfont $w_1$}
  \put(172.77,8.12){\fontsize{11.38}{13.66}\selectfont $w_2$}
  \put(146.04,8.12){\fontsize{11.38}{13.66}\selectfont $w_3$}
  \put(119.30,8.12){\fontsize{11.38}{13.66}\selectfont $w_4$}
  \put(89.22,8.12){\fontsize{11.38}{13.66}\selectfont $w_5$}
  \put(62.48,8.12){\fontsize{11.38}{13.66}\selectfont $w_6$}
  \put(35.75,8.12){\fontsize{11.38}{13.66}\selectfont $w_7$}
  \put(5.67,8.12){\fontsize{11.38}{13.66}\selectfont $w_8$}
  \end{picture}%
\caption{A tight instance of $27$ vertices,
	in which the $3$-path partition ${\cal Q}$ produced by the algorithm {\sc Approx3} contains twelve $2$-paths and one $3$-path (solid edges)
	and an optimal $3$-path partition ${\cal Q}^*$ contains nine $3$-paths (dashed edges).
	The six edges $(u_0, u_1)$, $(u_7, u_8)$, $(v_0, v_1)$, $(v_7, v_8)$, $(w_0, w_1)$, $(w_7, w_8)$ are in $\mcQ_2$ and $\mcQ^*_3$,
	shown in both solid and dashed (they are drawn overlapping).
	Note that the two edges $(u_4, v_4), (v_4, w_4)$ of $\mcQ_3$ are irrelevant edges,
	and there is no augmenting path with respect to $\mcQ$.
	One sees that $|\mcQ^*_{3,3}| = 6 = 2 |\mcQ^*_3|/3$, suggesting the performance analysis for the algorithm {\sc Approx3} is tight.\label{fig11}}
\end{figure}
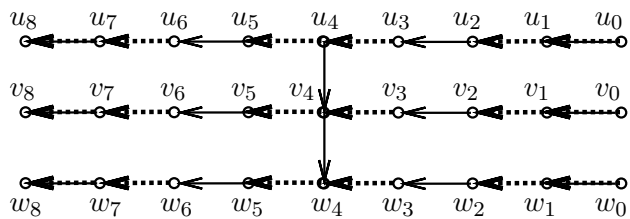

\section{Conclusions}\label{sec3}
We studied the {\sc $k$PP} problem on directed graphs, which seemingly escaped from the literature.
We proposed a novel concept of augmenting path to design a first $k/2$-approximation algorithm for the problem,
which is iterative and in each iteration it seeks to reduce the number of singletons until impossible.
When $k \ge 7$, we were able to design an improved $(k+2)/3$-approximation algorithm,
starting with the maximum path-cycle cover in the graph to carefully eliminate the $2$-cycles.
Certainly, this is also a $(k+2)/3$-approximation algorithm for the special case where the given graph is undirected,
improving the previously best approximation ratio of $k/2$~\cite{CGL19b}.

When $k = 3$, we defined the second kind of alternating and augmenting paths to reduce the number of $2$-paths
and presented an improved $13/9$-approximation algorithm.

See Table~\ref{tab01} for the summarized approximation results as of today.
Designing better approximation algorithms for $k$PP, in any listed case, is certainly interesting,
in particular, for $3$PP on directed graphs. 
Above all, an $o(k)$-approximation algorithm for $k$PP would be exciting.

On the other hand, when $k$ is part of the input, the $k$-path partition problem is APX-hard and can not be approximated within ratio $2$.
It would be interesting to know whether the {\sc $k$PP} problem is APX-hard, for any fixed $k \ge 3$,
and to see some non-trivial lower bounds on the approximation ratios.

\section*{Declarations}
\paragraph*{Data availability.}
Not applicable

\paragraph*{Interests.}
The authors declare that they have no known competing financial interests or personal relationships
that could have appeared to influence the work reported in this paper.



\newpage
\appendix
\section{A detailed analysis for the first $k/2$-approximation for $k$PP}\label{secA}
The performance analysis of the algorithm {\sc Approx1} for the $k$PP problem on directed graphs
can be regarded as a generalization of the undirected counterpart, yet nontrivial.
The design of the algorithm is presented in Subsection~\ref{sec21};
Lemmas~\ref{lemma08} and \ref{lemma09}, and the proof of Theorem~\ref{thm01} are include here.

\begin{lemma}
\label{lemma08}
Given a $k$-path partition $\mcQ$ in the directed graph $G = (V, E)$,
determining whether or not there exists an augmenting path with respect to $\mcQ$, and if so finding one such path, can be done in $O(m)$ time.
\end{lemma}
\begin{proof}
A single BFS traversal is sufficient since when two alternating paths starting with two distinct singletons meet,
they share everything from that point on and no re-exploration is necessary.
\end{proof}

\begin{lemma}
\label{lemma09}
Given a $k$-path partition $\mcQ$ in the directed graph $G = (V, E)$,
if there exists an augmenting path with respect to $\mcQ$,
then $\mcQ$ can be transferred into another $k$-path partition with at least one less singleton in $O(n)$ time.
\end{lemma}
\begin{proof}
Let $P$ denote the augmenting path with respect to $\mcQ$, and $s$ denote its starting vertex which is a singleton.
Let $(u, v)$ denote the free edge which ends $P$.

The prefix of $P$ from $s$ to $u$, denoted as $P(s, u)$, is an even length alternating path.
Replacing the matched edges of $\mcQ$ on $P(s, u)$ by the free edges on $P(s, u)$
transfers $\mcQ$ into another $k$-path partition $\mcQ'$ with exactly the same number of $i$-paths, for each $i = 1, 2, \ldots, k$, and additionally,
in $\mcQ'$ the vertex $u$ becomes a singleton and,
without loss of generality, $v$ is the $j$-th vertex $v_j$, where $j \ne 2$, on an $\ell$-path $v_1$-$v_2$-$\cdots$-$v_\ell$ of $\mcQ'$,
see Figure~\ref{fig01a} for illustrations.

It follows that, when $\ell \le 2$,
$j = 1$ and adding the free edge $(u, v_1)$ to $\mcQ'$ merges $u$ and the $\ell$-path $v_1$-$v_2$-$\cdots$-$v_\ell$ into an $(\ell+1)$-path
(see Figure~\ref{fig01a} for an illustration).
When $\ell \ge 3$ and $j = 1$,
adding the free edge $(u, v_1)$ to while removing the edge $(v_1, v_2)$ from $\mcQ'$ transfers
the singleton $u$ and the $\ell$-path $v_1$-$v_2$-$\cdots$-$v_\ell$ into a $2$-path $u$-$v_1$ and an $(\ell-1)$-path $v_2$-$\cdots$-$v_\ell$.
When $\ell \ge 3$ and $j \ge 3$,
adding the free edge $(u, v_j)$ to while removing the edge $(v_{j-1}, v_j)$ from $\mcQ'$ transfers
the singleton $u$ and the $\ell$-path $v_1$-$v_2$-$\cdots$-$v_\ell$ into a $(j-1)$-path $v_1$-$v_2$-$\cdots$-$v_{j-1}$
and an $(\ell-j+2)$-path $u$-$v_j$-$\cdots$-$v_\ell$.
Either way, the resulting $k$-path partition contains at least one less singleton.
One clearly sees that the transferring process takes $O(n)$ time by walking through the augmenting path $P$ once.
\end{proof}

\noindent
\begin{proof}
(of Theorem~\ref{thm01})
We first show that the $k$-path partition $\mcQ$ returned by the algorithm {\sc Approx1}
achieves the minimum number of singletons among all $k$-path partitions.
The proof is done by constructing a mapping from the singletons in $\mcQ$ to the singletons of any other $k$-path partition $\mcQ'$,
using the alternating paths.
First, if $s$ is a singleton in both $\mcQ$ and $\mcQ'$, then $s$ is mapped to $s$.
Below we consider $s$ being a singleton in $\mcQ$ but not in $\mcQ'$.
We assume w.l.o.g. that the edge of $\mcQ'$ incident at $s$ leaves $s$, that is, $(s, v)$.
So $(s, v)$ is a free edge with respect to $\mcQ$.

Due to the non-existence of an augmenting path,
we conclude that $v$ is the second vertex $v_2$ of an $\ell$-path $v_1$-$v_2$-$\cdots$-$v_\ell$ of $\mcQ$
(see Figure~\ref{fig01b} for an illustration).
Since the free edge $(s, v_2)$ is in $\mcQ'$, the matched edge $(v_1, v_2)$ cannot be in $\mcQ'$ and it can be ``discovered'' only by the free edge $(s, v_2)$ of $\mcQ'$.

We distinguish two cases:
In Case 1, $v_1$ is a singleton in $\mcQ'$.
Then, adding the free edge $(s, v_2)$ to and removing the matched edge $(v_1, v_2)$ from $\mcQ$ will transfer $\mcQ$ into another $k$-path partition in which
$s$ is no longer a singleton but $v_1$ becomes a singleton.
In this sense, we say that the alternating path {\em maps} the singleton $s$ of $\mcQ$ to the singleton $v_1$ of $\mcQ'$.

In Case 2, $v_1$ is not a singleton in $\mcQ'$.
The edge of $\mcQ'$ incident at $v_1$, either entering or leaving $v_1$, is a free edge with respect to $\mcQ$.
We assume w.l.o.g. that this edge leaves $v_1$, that is, $(v_1, w)$.
One sees that $v_1$ takes up the same role as the singleton $s$ in the above argument,
and again by the algorithm $w$ has to be the second vertex $w_2$ of some $\ell'$-path $w_1$-$w_2$-$\cdots$-$w_{\ell'}$ of $\mcQ$.
Note that similarly the matched edge $(w_1, w_2)$ cannot be in $\mcQ'$ and it can be ``discovered'' only by the free edge $(v_1, w_2)$ of $\mcQ'$.
We then repeat the above discussion on $w_1$, either to have an alternating path mapping the singleton $s$ of $\mcQ$ to the singleton $w_1$ of $\mcQ'$,
or to use the free edge in $\mcQ'$ that is incident at $w_1$ to extend the alternating path.
({\it Symmetrically, if the free edge of $\mcQ'$ enters $v_1$, that is, $(w, v_1)$, 
	then $w$ has to be the second last vertex $w_{\ell'-1}$ of some $\ell'$-path $w_1$-$w_2$-$\cdots$-$w_{\ell'}$ of $\mcQ$.
	Similarly the matched edge $(w_{\ell'-1}, w_{\ell'})$ cannot be in $\mcQ'$ and it can be ``discovered'' only by the free edge $(w_{\ell'-1}, v_1)$ of $\mcQ'$.
	Consequently, we either have an alternating path mapping the singleton $s$ of $\mcQ$ to the singleton $w_{\ell'}$ of $\mcQ'$,
	or use the free edge in $\mcQ'$ that is incident at $w_{\ell'}$ to extend the alternating path.
	See Figure~\ref{fig01d} for an illustration.})
Due to the finite order of the graph $G$, at the end we will have an alternating path mapping the singleton $s$ of $\mcQ$ to a singleton of $\mcQ'$.

Using the fact that there is at most one edge of $\mcQ$ ($\mcQ'$, respectively) leaving each vertex 
and at most one edge of $\mcQ$ ($\mcQ'$, respectively) entering each vertex,
a singleton of $\mcQ'$ is not mapped by multiple singletons of $\mcQ$.
Since every singleton of $\mcQ$ is mapped to a singleton of $\mcQ'$,
we conclude that the number of singletons in $\mcQ$ is no greater than the number of singletons in $\mcQ'$.

Let $\mcQ^*$ denote an optimal $k$-path partition that minimizes the number of paths.
Also, let $\mcQ^*_i$ ($\mcQ_i$, respectively) denote the sub-collection of all the $i$-paths of $\mcQ^*$ ($\mcQ$, respectively), for $i = 1, 2, \ldots, k$.
It follows that
\[
\sum_{i=1}^k i |\mcQ_i| = n = \sum_{i=1}^k i |\mcQ^*_i| \mbox{ and } |\mcQ_1| \le |\mcQ^*_1|.
\]
Adding them together we have
\[
2 |\mcQ_1| + 2 |\mcQ_2| + 3 |\mcQ_3| + \cdots + k |\mcQ_k| \le 2 |\mcQ^*_1| + 2 |\mcQ^*_2| + 3 |\mcQ^*_3| + \cdots + k |\mcQ^*_k|,
\]
which leads to $2 |\mcQ| \le k |\mcQ^*|$ and thus proves the theorem.
\end{proof}

\end{document}